\documentclass[journal=jpccck,manuscript=article,layout=twocolumn]{achemso}

\usepackage{chemformula} 
\usepackage[T1]{fontenc} 

\usepackage{graphicx}
\usepackage{amsmath}
\usepackage{amssymb}
\usepackage{amsthm}
\usepackage{tabularx}
\usepackage{multirow}

\SectionNumbersOn

\newcommand{\dirimg}{./}

\title{Formation and dynamics of small polarons on the rutile TiO$_2$(110) surface}

\author{Michele Reticcioli}
\affiliation{University of Vienna, Faculty of Physics and Center for Computational Materials Science, Vienna, Austria}
\author{Martin Setvin}
\affiliation{Institute of Applied Physics, Technische Universit\"at Wien, Vienna, Austria}
\author{Michael Schmid} 
\affiliation{Institute of Applied Physics, Technische Universit\"at Wien, Vienna, Austria}
\author{Ulrike Diebold}
\affiliation{Institute of Applied Physics, Technische Universit\"at Wien, Vienna, Austria}
\author{Cesare Franchini}
\email{cesare.franchini@univie.ac.at}
\affiliation{University of Vienna, Faculty of Physics and Center for Computational Materials Science, Vienna, Austria}

\abbreviations{IR,NMR,UV}

\begin{document}

\begin{tocentry}
        \includegraphics[width=1\columnwidth,clip=true]{\dirimg 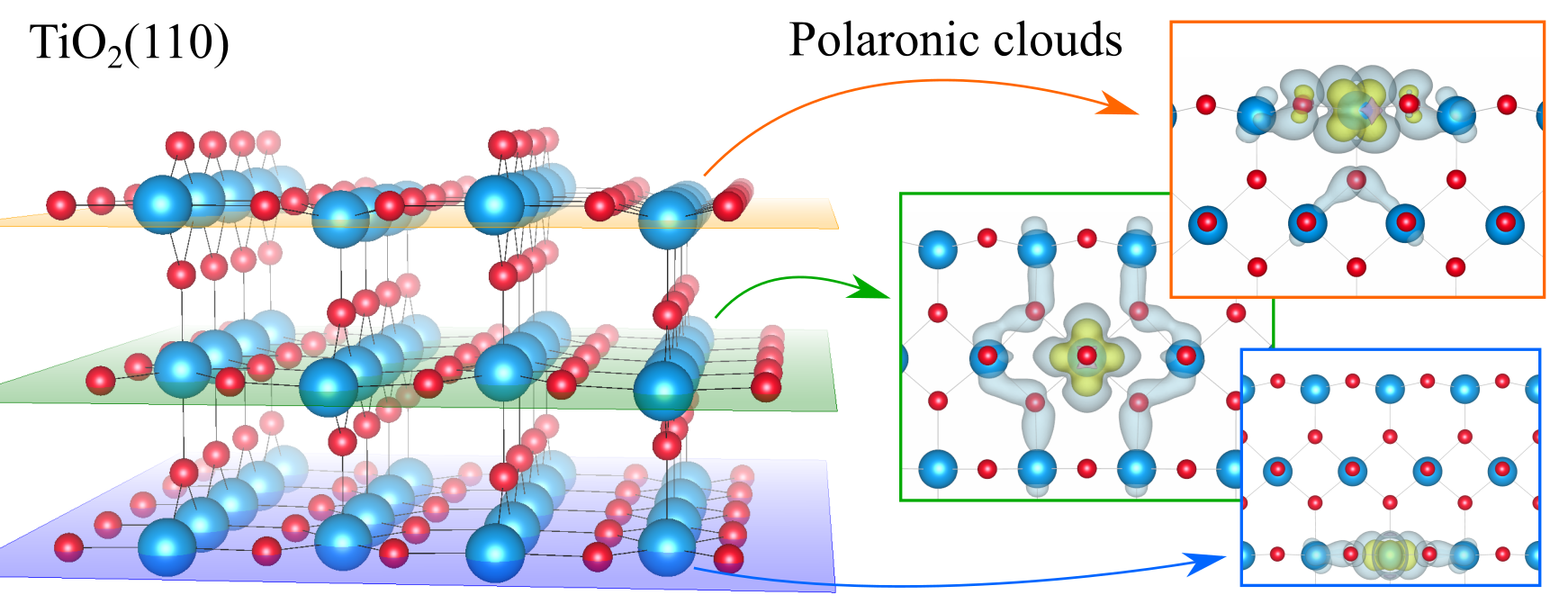}
\end{tocentry}

\begin{abstract} 
Charge trapping and formation of polarons is a pervasive phenomenon in transition metal oxide compounds, in particular at the surface, affecting fundamental physical properties and functionalities of the hosting materials.
Here we investigate via first-principle techniques the formation and dynamics of small polarons on the reduced surface of titanium dioxide, an archetypal system for polarons, for a wide range of oxygen vacancy concentrations.
We report how the essential features of polarons can be adequately accounted in terms of few quantities: the local structural and chemical environment, the attractive interaction between negatively charged polarons and positively charged oxygen vacancies, and the spin-dependent polaron-polaron Coulomb repulsion.
We combined molecular dynamics simulations on realistic samples derived from experimental observations with simplified static models, aiming to disentangle the various variables at play.
We find that depending on the specific trapping site, different types of polarons can be formed, with distinct orbital symmetries and different degree of localization and structural distortion.
The energetically most stable polaron site is the subsurface Ti atom below the undercoordinated surface Ti atom, owing to a small energy cost to distort the lattice and a suitable electrostatic potential.
Polaron-polaron repulsion and polaron-vacancy attraction determine the spatial distribution of polarons as well as the energy of the polaronic in-gap state.
In the range of experimentally reachable oxygen vacancy concentrations the calculated data are in excellent agreement with observations, thus validating the overall interpretation.
\end{abstract}

\section{Introduction}
Excess electrons present in transition metal oxides can locally couple to lattice distortions and form small polarons~\cite{Stoneham1989,Shluger1993,Stoneham2007}.
The presence of such a localized charge affects the physical and chemical properties of the hosting material, with a local alteration of the bond lengths, a change of the formal valence at the specific polaronic site, and the emergence of a characteristic peak localized in the gap region~\cite{Nagels1963,Verdi2017,Sezen2015a,Freytag2016}.
Small polarons play a decisive role in electron transport~\cite{Crevecoeur1970,Moser2013}, optical absorption, and chemical reactivity, and have crucial implications in other diverse phenomena including high-$T_{\rm c}$ superconductivity~\cite{Salje}, colossal magnetoresistance~\cite{Teresa,Ronnow}, thermoelectricity~\cite{Wang2014}, photoemission~\cite{Cortecchia2017}, and photochemistry~\cite{Linsebigler}.
Here we focus on the (110) surface of rutile TiO$_2$, TiO$_2$(110), a highly studied oxide surface~\cite{Diebold2003} for which the presence of small polarons was predicted by different computational approaches~\cite{DiValentin2006,Deak2012b,Berger2015b,Yan2015} and confirmed by several experiments~\cite{Kruger2008,Yang2013,Setvin2014}.
Excess electrons are found to originate mostly from surface defects, such as oxygen vacancies, Nb impurities, hydroxyl groups and interstitial titanium atoms~\cite{Shibuya2014,Setvin2014,Mao2013,Finazzi2009}.
Formation of polarons is particularly favorable at Ti$^{4+}$ sites in the near-surface region~\cite{Deskins2011}, with consequent effects on the adsorption~\cite{Cao2017a} and on the stability of the surface.
At high reduction states, the strong polaronic repulsion was found to drive the surface from a (1$\times$1) termination to a (1$\times$2) reconstruction~\cite{Reticcioli2017c}.
Therefore, it is of fundamental importance to understand the mechanisms behind the formation of polarons, in order to possibly optimize or tune existing applications~\cite{Carneiro2017} and propose novel functionalities.

The study of the polaronic properties is revealed by the simultaneous presence of multiple interactions.
First, an isolated polaron is coupled with local atomic distortions that are expected to depend on the specific structural symmetry of localization sites~\cite{Setvin2014b}, and are generally in the range 0.02-0.15~\AA.
Second, being localized negatively-charged quasi-particles, small polarons strongly repel each other.
Finally, defects often present in this system, such as surface oxygen vacancies, could interact with the polarons electronically and also via lattice distortions induced by the defect itself.
Here, to extricate the importance and effect of the distinct contributions we combine extensive molecular dynamic calculations on realistic structural models for a wide range of oxygen vacancy concentration with idealized static models based on specific and judicious choices regarding the concentration of excess electrons, the polaron distribution and  the presence or absence of oxygen vacancies. For instance, tuning the amount of excess electrons in the pristine slab enables the study of the properties of isolated polarons and of the polaron-polaron interaction in absence of oxygen vacancies, a situation which is difficult to achieve in realistic samples.
This computational procedure allows us to unravel the mechanisms at play and describe fundamental polaronic properties.
We found that the optimal polaronic distribution and mobility is the result of the balance between several -- and to some extent antagonistic -- factors:
(i) minimization of the distance between polaron and oxygen vacancies at any concentration, (ii) maximization of the polaron-polaron distance and (iii) propensity for polaron formation in the sub-surface layer, rather than in the bulk or surface regions. 
Our results complement our previous analysis on the polaron-induced structural instability of the TiO$_2$(110) surface~\cite{Reticcioli2017c}, are validated by comparison with existing experiments, and are consistent with the few previous studies dealing with the hopping mobility of polarons in TiO$_2$~\cite{Kowalski2010,Deskins2007,Setvin2014}.

\section{Methods}
\label{methods}

\begin{figure*}[t]
    \begin{center}
        \includegraphics[width=1.9\columnwidth,clip=true]{\dirimg 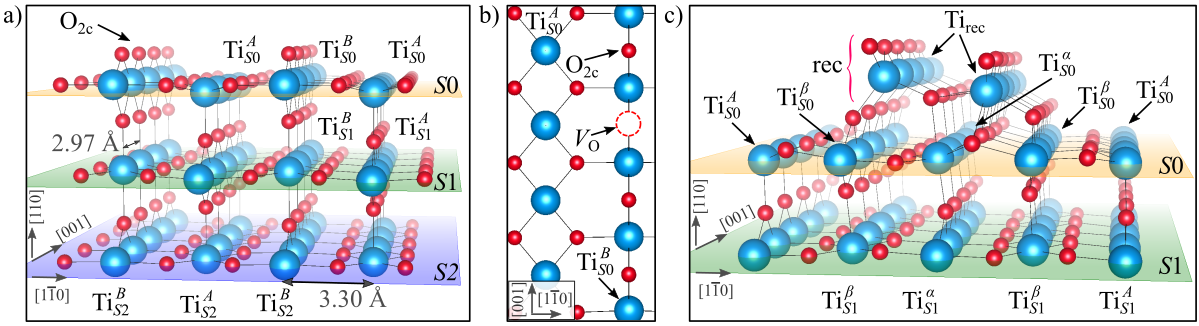}
    \end{center}
\caption{TiO$_2$(110) surface structures. Front view (a) of the pristine (1$\times$1) phase, sketched top view (b) of the reduced (1$\times$1) phase and front view (c) of the (1$\times$2) reconstruction. Reported distances refer to the pristine (1$\times$1) surface.
}
\label{fig:structures}
\end{figure*}

We addressed the study of polarons in rutile TiO$_2$(110) surface in the framework of density functional theory (DFT), by using the Vienna \emph{ab initio} simulation package (VASP)~\cite{Kresse1996a,Kresse1996}.
We adopted the generalized gradient approximation (GGA) within the Perdew, Burke, and Ernzerhof parametrization~\cite{Perdew1996} with the inclusion of an on-site effective $U$~\cite{Dudarev1998} of $3.9$~eV on the $d$ orbitals of the Ti atoms~\cite{Wang2017}, previously determined by constrained-random-phase-approximation calculations in bulk rutile~\cite{Setvin2014}.
We modeled the (1$\times$1) reduced rutile surfaces with an asymmetric slab containing five Ti layers in a large two-dimensional 9$\times$2 unit cell [see Fig.~\ref{fig:structures}(a,b)].
The (1$\times$2) reconstructed phase was constructed according to the Ti$_2$O$_3$ model~\cite{Onishi1994,Wang2014a,Mochizuki2016} by placing reconstructed asymmetric rows on top of the five layers of the (1$\times$1) phase [see Fig.~\ref{fig:structures}(c)].
The bottom two layers were kept fixed whereas all the other atomic sites where relaxed using standard convergence criteria with a plane-wave energy cutoff of 300~eV, and using the $\Gamma$ point only for the integration in the reciprocal space.
For the (1$\times$1) phase, up to 9~oxygen vacancies ($V_{\rm O}$'s) were homogeneously included in the top layer at different concentrations ($c_{V_{\rm O}}=5.6\%,11.1\%,16.7\%,22.2\%,33.3\%,38.9\%,50.0\%$)~\cite{Cheng2009,Li2015}.
Each V$_{\rm O}$ supplies two excess electrons, eligible to form two polarons~\cite{Setvin2014}.
The $50\%$ deviation from the stoichiometry of the reconstructed surface provides the slab with two excess electrons per (1$\times$2) unit cell~\cite{Reticcioli2017c}.

We performed first-principle molecular dynamics (FPMD)~\cite{Kresse1993} on these slabs, in order to analyze the hopping behavior of polarons~\cite{Hao2015a}.
The FPMD was conducted at a simulation temperature of 700~K with a timestep of 1~fs for at least 10~ps (17~ps for the $c_{V_{\rm O}}=5.6\%$ slab), with a lower energy cutoff of 250~eV.
A statistical analysis was performed on the FPMD results.
The hopping behavior of polarons was analyzed as function of the distance from the surface by determining the number of occurrences of charge trapping in each layer during every FPMD run.
We computed also the polaron-polaron site correlation function $S_{\rm pol-pol}$, defined as the distribution of the site distance $i$ along [001] between two polarons at a given time-step $t$, averaged over the complete FPMD interval $\tau$:
\begin{equation}
S_{\rm pol-pol}(i)= \frac 1 {N} \frac 1 {\tau} \sum_t \sum_{j} \rho_j(t) \rho_{j+i}(t)~,
\end{equation}
where $N$ is the number of Ti sites, and $\rho_j(t)$ indicates the polaronic site density at time $t$, and it is equal to 1 for the $j$-th Ti site hosting a polaron, and 0 otherwise.
Analogously, we computed the polaron-polaron $R_{\rm pol-pol}$ and vacancy-polaron $R_{\rm V_{O}-pol}$ radial correlation functions as a function of the distance $r$:
\begin{equation}
\begin{split}
      R_{\rm pol-pol}(r)&=  \frac 1 {\tau} \sum_t \sum_{(q,p)} \delta(|\mathbf{r}_q-\mathbf{r}_p|,r,t)~,\\
R_{\rm V_{O}-pol}(r)&=  \frac 1 {\tau} \sum_t \sum_{(V_{\rm O},p)} \delta(|\mathbf{r}_{V_{\rm O}}-\mathbf{r}_p|,r,t)~,
\end{split}
\end{equation}
where the variables $\delta(|\mathbf{r}_{V_{\rm O}}-\mathbf{r}_q|,r,t)$ and $\delta(|\mathbf{r}_q-\mathbf{r}_p|,r,t)$ assume the value 1 if, at time $t$, the polaron $p$ is at distance $r$ from the $V_{\rm O}$ at position $\mathbf{r}_{V_{\rm O}}$ or from the polaron $q$ at position $\mathbf{r}_q$, respectively, and are 0 otherwise.

Furthermore, we performed another set of DFT+$U$ calculations considering the approximately 200 inequivalent polaronic configurations obtained from each FPMD simulation.
In this set of post-FPMD DFT+$U$ calculations, the structures corresponding to the various inequivalent polaronic configurations were further relaxed at $T=0$~K.
This allows us to calculate the total energy $E^{\rm loc}_{\rm relax}$ of each configuration and the polaron formation energy $E_{\rm POL}$ as
\begin{equation}
 E_{\rm POL}= E^{\rm loc}_{\rm relax} - E^{\rm deloc}_{\rm relax}~,
\end{equation}
where $E^{\rm deloc}_{\rm relax}$ is the ground state energy of the system forced to have all the electrons delocalized.
The delocalized solution was achieved by performing non spin-polarized calculations.
The stability of a polaron, quantified by $E_{\rm POL}$, is the result of the competition between the structural cost needed to distort the lattice in order to accommodate polarons ($E_{\rm ST}$), and the electronic energy gained by localizing the electron in the distorted lattice ($E_{\rm EL}$)~\cite{Setvin2014}:
\begin{equation}
 E_{\rm POL}= E_{\rm EL} + E_{\rm ST}~,
\end{equation}
where $E_{\rm ST}$ is defined as
\begin{equation}
 E_{\rm ST}= E^{\rm deloc}_{\rm constr} - E^{\rm deloc}_{\rm relax}~,
\end{equation}
with $E^{\rm deloc}_{\rm constr}$ being the energy of the system forced to have only delocalized electrons and constrained into the structure of the system hosting polarons.

Besides the FPMD and post-FPMD calculations, we performed a set of static DFT+$U$ calculations for selected model structures in order to investigate individually the key quantities driving the polaron formation at the (1$\times$1) surface. 
To this aim, we altered the charged-neutral state of the system, with and without oxygen vacancies, by modeling the following charged configurations:
\begin{enumerate}
 \item[(i)] No oxygen vacancies and one excess electron ($-1$ charged system).
 This setup allows to study individual polarons with no perturbations coming from $V_{\rm O}$.
 \item[(ii)] One oxygen vacancy with only one excess electron (i.e. one of the two excess electrons provided by the $V_{\rm O}$ is neutralized by the manual addition of one extra hole) resulting in a $+1$ charged system.
This enables us to inspect directly the effect of $V_{\rm O}$ on the polaronic properties.
 \item[(iii)] No oxygen vacancies and two excess electrons ($-2$ charged system), to study polaron-polaron interactions.
\end{enumerate}
In addition to the 9$\times$2 cells, we also built thicker 3$\times$2 and 5$\times$2 slabs containing 8 rather than 5 Ti layers (2 of which were kept fixed at the bulk positions) in order to study the properties of polaron formation as a function of the depth.
We calculated the relevant quantities for polaron formation (such as $E_{\rm POL}$, $E_{\rm ST}$, $E_{\rm EL}$) as a function of the position of a polaron in the cell.
The charged systems are automatically neutralized by a homogeneous-background charge.
We notice that, since the polaronic energies are defined as differences (between localized and delocalized solutions at constant number of electrons), adopting charged systems does not substantially alter the results, regardless the surface extension of the slabs, as we tested for the 3$\times$2, 5$\times$2, and 9$\times$2 slabs.

The excess electrons were localized in specific Ti sites of the cell according to the following strategy, which includes three consequential DFT+$U$ calculations~\cite{Deskins2011,Hao2015a}:
\begin{enumerate}
 \item Vanadium chemical substitution at the Ti site(s) chosen to host the polaron(s).
 This structure is relaxed by a DFT+$U$ calculation, which typically yields to a strong distortion of the lattice around the V site(s).
 \item V atom(s) are replaced by the original Ti atom(s), and a new relaxation is performed by selectively imposing a larger value of $U$ of $9.9$~eV to facilitate the localization of the excess electron(s) at the chosen Ti atom(s).
 \item Final relaxation using the proper $U$ value of $3.9$~eV for all Ti atoms, including the polaronic site.
A practical note: this final step was started by using the wavefunction of step (2) to facilitate the localization of the excess electron at the chosen site.
By using random coefficients for the initial wavefunctions, the calculation could lead to a polaron localized in a different Ti site or to a delocalized solution (excess electron in the conduction band)~\cite{Papageorgiou2010a,Deskins2009,Meredig2010a,Morgan2009}.
\end{enumerate}

The slabs adopted for the static model were also used to estimate the energy barrier for polaron hopping between two Ti sites.
To this aim, we built intermediate polaronic structures by linearly interpolating the initial and final polaronic structures~\cite{Janotti2013a}.
An electronic self consistent loop at fixed geometry was performed in order to calculate the total energy at each step.

We also analyzed the contribution of the electrostatic potential on the polaron formation and dynamics by inspecting the volume-averaged electrostatic potential energy for the electrons ($E_{\rm pot}$) at each atomic site (this was computed also in an additional setup, that is the $+2$ charged state: one $V_{\rm O}$ plus two extra holes neutralizing both the excess electrons).
The density of states (DOS) and the distribution of the polaronic charge density in real space were obtained using a large plane-wave energy cutoff of 700~eV.

In the following, we refer to Fig.~\ref{fig:structures} for the notation of the atomic sites in TiO$_2$(110) surfaces.
In particular, the $A$ sites are the five-fold coordinated Ti atoms on the top layer $S0$ and all the octahedrally coordinated Ti atoms at deeper layers ($S1$=subsurface, $S2$=sub-subsurface, etc\dots) below the Ti$^A_{S0}$ row.
The $B$ sites are the Ti atoms bonded to the two-fold coordinated O$_{2c}$ atoms and/or $V_{\rm O}$ in the $S0$ layer and all the octahedrally coordinated Ti atoms at deeper layers below the Ti$^B_{S0}$ row.
For the (1$\times$2) reconstruction, the Ti sites below the reconstructed rows are named with Greek symbols, with $\alpha$ and $\beta$ replacing $A$ and $B$, respectively.

\section{FPMD Results}
\label{results}

\begin{figure*}[t]
    \begin{center}
        \includegraphics[width=1.9\columnwidth,clip=true]{\dirimg 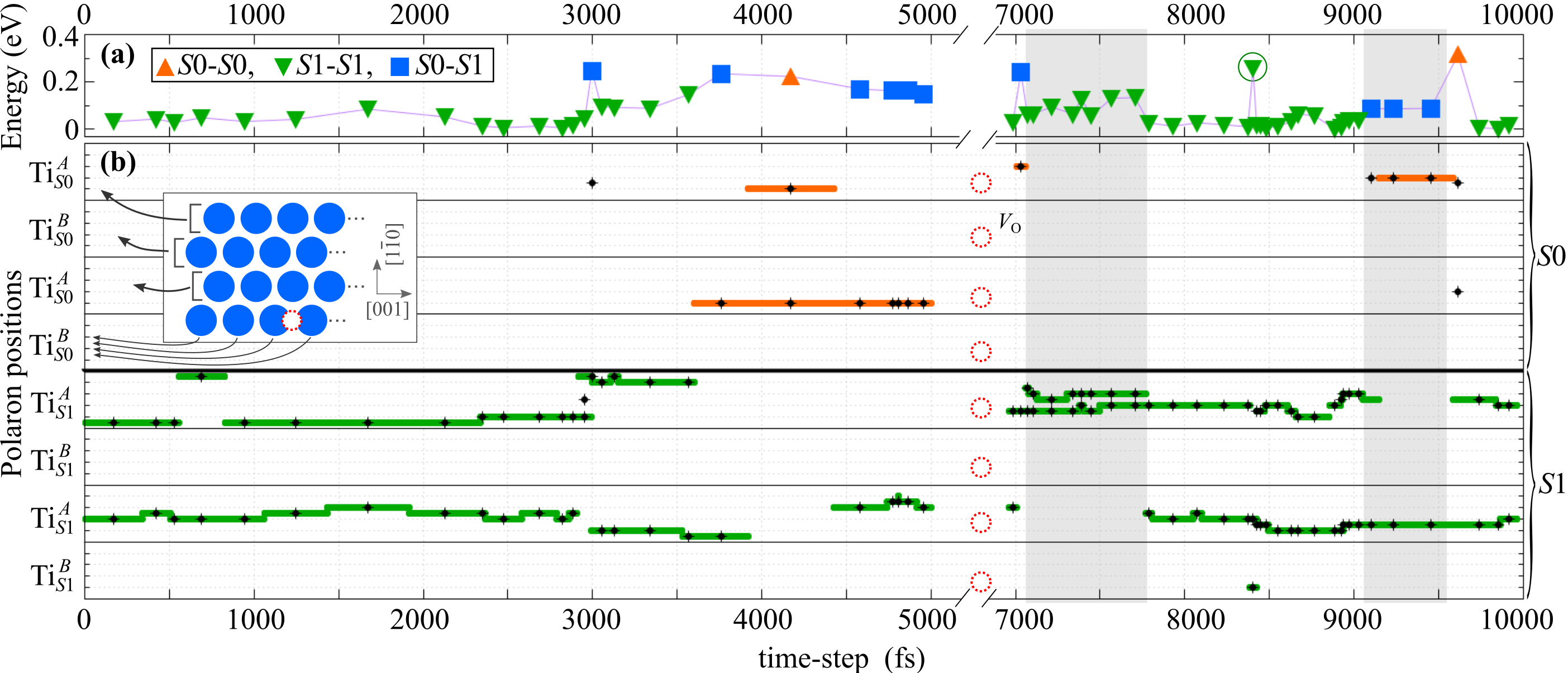}
    \end{center}
\caption{FPMD polaron hopping. A representative part of the results for the (1$\times$1) surface with one oxygen vacancy in the 9$\times$2 slab ($c_{V_{\rm O}}=5.6\%$) is shown. The gray rectangles highlight data discussed in detail in the main text. Panel (a): $T=0$ DFT+$U$ energies (based on the structures obtained by FPMD at $700$~K and fully relaxed at $T=0$~K). The most stable polaronic configuration is taken as a reference. The up-pointing triangles, down-pointing triangles and squares refer to configurations with the two excess electrons in the slab localized both at the $S0$ layer, both at the $S1$ layer and one polaron per $S0$ and $S1$ layer, respectively. The circled value indicates the case of a polaronic Ti$^B_{S1}$ site. Panel (b): polaron positions at every FPMD step. The $y$-axis indicates the polaron positions along each Ti$^A$ and Ti$^B$ [001] row (as sketched in the inset). The thick lines indicates the polaron positions as obtained by the FPMD runs at $T=700$~K, whereas crosshairs are the corresponding post-FPMD polaron configurations obtained at $T=0$~K. The position of $V_{\rm O}$ projected along the various [001] rows is shown by dashed circles.
}
\label{fig:hopping}
\end{figure*}

\begin{figure}[t]
    \begin{center}
        \includegraphics[width=0.9\columnwidth,clip=true]{\dirimg 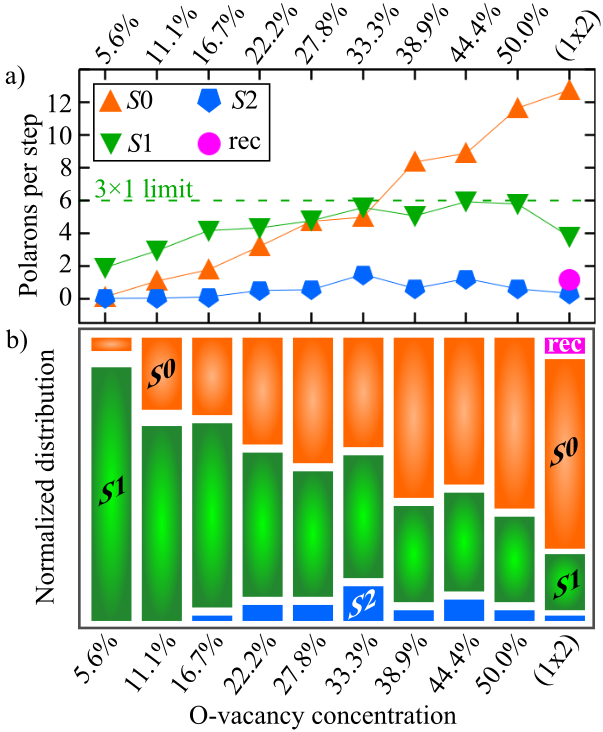}
    \end{center}
\caption{FPMD layer-resolved statistical analysis. Panel (a): Average number of polarons at the various layers per FPMD time step. The dashed line represents the number of polarons (\textit{i.e.} six polarons) sustained by the 3$\times$1 pattern in our 9$\times$2 slab. Panel (b): Overall occurrences of polaron formation at the various layers. For the 1$\times$1 phase, the histogram bars represent from top to bottom the $S$0, $S$1 and eventually $S$2 layers, while for the 1$\times$2 phase the topmost bar refers to the reconstructed layer.
}
\label{fig:histo}
\end{figure}

\begin{figure*}[t]
        \includegraphics[width=2\columnwidth,clip=true]{\dirimg 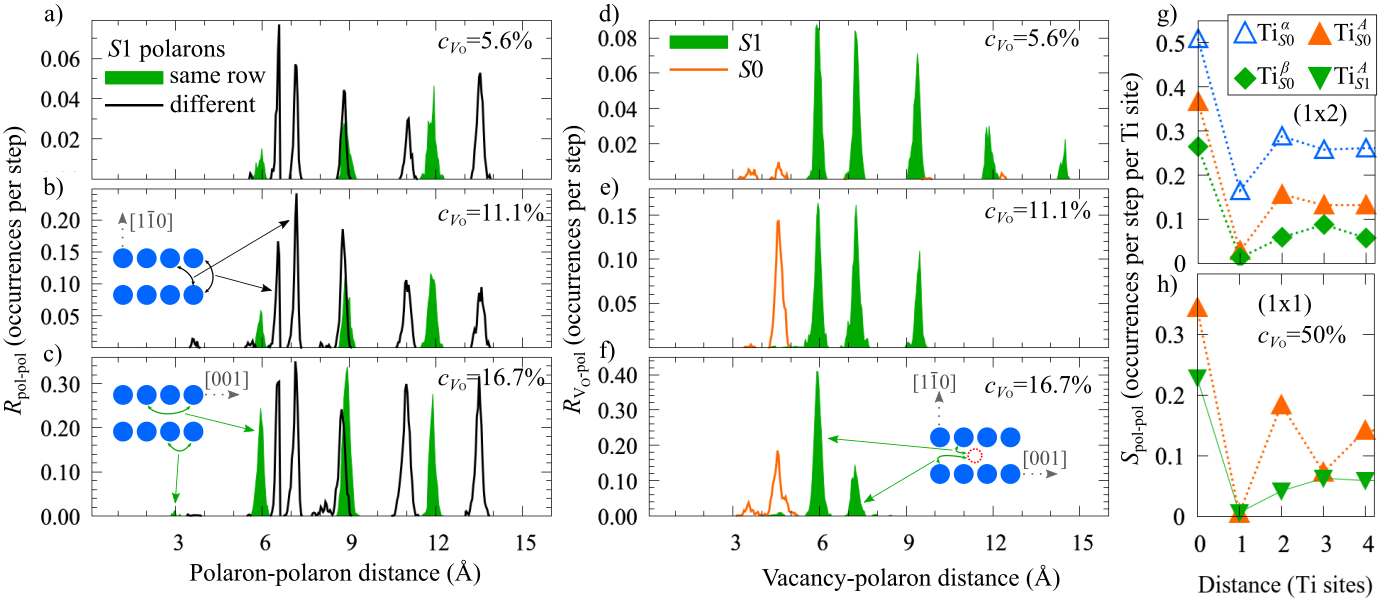}
        \caption{FPMD correlation. The radial distribution functions of the correlation between polarons (a,b,c) and between a polaron and the nearest oxygen vacancy (d,e,f) are shown for the systems with $5.6\%$, $11.1\%$ and $16.7\%$ oxygen vacancy concentration. Filled and empty curves refer, in panels (a,b,c), to the correlation between two $S1$ polarons in the same and different [001] rows, respectively, and, in panels (d,e,f), to the correlation between the oxygen vacancy and the $S1$ and $S0$ polaron, respectively.
The positions of the Ti$_{S1}^A$ (filled circles) and $V_{\rm O}$ (empty circle) sites are sketched.
The site-resolved polaron-polaron correlation function along [001] is shown for the (1$\times$2) and 50\% reduced (1$\times$1) cases (g,h). Here, the site-resolved correlation at null distance indicates the density of polarons per step on the various types of Ti atoms.
}
\label{fig:corr}
\end{figure*}

We performed FPMD calculations to analyze the polaron hopping at the reduced and reconstructed rutile TiO$_2$(110) surfaces.
Figure~\ref{fig:hopping} shows the results for the lowest $V_{\rm O}$ concentration, $c_{V_{\rm O}}=5.6\%$.
Here, two excess electrons originate from the oxygen vacancy in the slab, and form two polarons.
These polarons are initially localized at two sub-surface Ti$_{S1}^A$ sites, and hop to different Ti sites in the $S0$ and $S1$ layers, in particular sub-surface  Ti$_{S1}^A$ sites and surface Ti$_{S0}^A$ sites~\cite{Kowalski2010, Setvin2014} [Fig.~\ref{fig:hopping}(b)].
The distinctly different configurations assumed by the polarons during the FPMD run were further analyzed by DFT relaxations at $T=0$~K.
The resulting total energies are shown in the top panel of Fig.~\ref{fig:hopping}(a).
The specific arrangement of small polarons plays a crucial role in determining the energy and stability of TiO$_2$(110).
According to DFT at $T=0$~K, the most stable configurations are those hosting both polarons at Ti$_{S1}^A$ sites located in adjacent Ti$_{S1}^A$ rows (in agreement with previous calculations~\cite{Deskins2011,AmoreBonapasta2009} and experiments~\cite{Kruger2008}).
Formation of two polarons in the same Ti$_{S1}^A$ row has a high energy cost, about 150~meV with respect to the most stable configuration, and is indeed a relatively rare event (see gray area around 7500~fs in Fig.~\ref{fig:hopping}).
Polarons hop easily between adjacent sites along the Ti$_{S1}^A$ row~\cite{Yan2015a}, which is quantified by an energy barrier lower than approximately $200$~meV, as we estimated by interpolation of the atomic positions.

Polaron formation at $S0$ sites (filled squares in Fig.~\ref{fig:hopping}(a)) is largely disfavored, leading to an energy increase of about 200~meV.
In this case, our calculations yield an energy barrier of approximately $350$~meV for a polaron hopping from a Ti$_{S1}^A$ to a Ti$_{S0}^A$ site.
However, depending on the specific location of polarons in $S0$, significant energy changes occur.
As an example, the gray area around 9000-9500~fs in Fig.~\ref{fig:hopping} highlights a $S0$-$S1$ configuration which is only 90~meV less stable than the optimal Ti$_{S1}^A$ arrangement.
This behavior can be traced back to the relative distance and interaction between the $S0$ and $S1$ polarons as well by their interaction with the $V_{\rm O}$.
This issue will be discussed in more detail later on.
At variance with charge trapping at Ti$_{S1}^A$ sites, polaron formation at Ti$_{S1}^B$ sites occurs very rarely~\cite{Kruger2008}, only one observed event during an entire FPMD run, and the energy of the corresponding configuration relaxed at $T=0$~K is comparable to the $S0$-configurations (circled triangle in Fig.~\ref{fig:hopping}(a)).

We notice that in a few cases, during the $T=0$ relaxations the polaronic configuration changes with respect to the one at $700$~K.
We recall that we report the polaronic energies in terms of the final configurations assumed at $T=0$~K, which were also used to determine the most favorable configuration at each $c_{V_{\rm O}}$ level~\cite{Reticcioli2017c}.

In order to quantitatively describe the polaronic hopping, we performed a statistical analysis on the polaronic configurations obtained in the FPMD run.
Figure~\ref{fig:histo} shows the layer-resolved polaronic distribution for all oxygen-vacancy concentrations considered.
At $c_{V_{\rm O}}=5.6\%$, polaron hopping occurs mainly among the Ti sites at the $S$1 layer [see Fig.~\ref{fig:histo}(b)].
For higher concentrations the polarons populate more often the $S0$ layer, and for $c_{V_{\rm O}}>16.7\%$, also the $S2$ layer becomes sporadically populated by polarons.
At large $c_{V_{\rm O}}$'s polaron localization in $S0$ becomes progressively predominant.
This trend is due to the balance between two opposite effects:
The ease to host polarons in the sub-surface layer, and in particular in Ti$^A_{S1}$ sites, (dominant at low $c_{V_{\rm O}}$'s) and the strong repulsion between nearby polarons (dominant at high $c_{V_{\rm O}}$'s)~\cite{Reticcioli2017c}.
We previously reported that the maximum density of polarons in $S1$ is 16.7\%, with an optimal arrangement involving 1 polaron every 3 Ti sites along the Ti$^A_{S1}$ rows and no polaron in Ti$^B_{S1}$~\cite{Reticcioli2017c}.
In the employed 9$\times$2 cell this leads to a maximum density of 6 polarons in the $S$1 layer, arranged in a 3$\times$1 pattern.
Indeed, Fig.~\ref{fig:histo}(a) shows that the polaron density in $S1$ does not exceed this limit. 
At larger $c_{V_{\rm O}}$ polarons prefer to populate $S0$ rather than undermining the 3$\times$1 favorable configuration in the Ti$^A_{S1}$ rows.
In contrast to Ti$^A_{S1}$ polarons, excess electrons in $S0$ reach higher densities in the Ti$^A_{S0}$ rows, and localize also at the Ti$^B_{S0}$ atoms near the $V_{\rm O}$.
The increasing polaron-polaron repulsion gradually weakens the stability of the surface and ultimately leads to a (1$\times$2)-$\rm Ti_2O_3$ structural reconstruction, which is able to accommodate a large amount of excess electrons at easily reducible Ti$_{S0}$ sites.
The details of this reconstruction were discussed in our previous work~\cite{Reticcioli2017c}.

Figure~\ref{fig:corr} collects information on the statistical analysis of the polaron-polaron and polaron-vacancy interactions, which are the key quantities for achieving a reliable picture of the formation and dynamics of polarons. 
The correlation function $R_{\rm pol-pol}$ displayed in Fig.~\ref{fig:corr}(a-c) shows that polarons in nearest-neighbor sites along the [001] Ti$^A_{S1}$ row (i.e. at an inter-polarons distance of 2.97~\AA) are extremely rare.
Instead, polarons prefer to maximize the distance between them as manifested by the strong peak at 9~\AA\ and 12~\AA, corresponding to a 4$\times$ and 3$\times$ periodicity in the 9-sites long Ti$^A_{S1}$ row at $c_{V_{\rm O}}=5.6\%$-$11.1\%$ and $c_{V_{\rm O}}=16.7\%$, respectively. 

The correlation between polarons lying at different [001] Ti$^A_{S1}$ rows is rather homogeneous with a larger probability to find polaron pairs at short distances, approximately corresponding to the inter-row distance of 6.6~\AA\ [Fig.~\ref{fig:corr}(a,b)].
This implies that inter-row polaron-polaron repulsion is essentially ineffective.
The nature of these peaks at short distances becomes clear by considering the contribution of the vacancy, which is summarized in Fig.~\ref{fig:corr}(d-f), in terms of the correlation function $R_{\rm V_{O}-pol}$.
These data show a polaron-${V_{\rm O}}$ attraction, which clearly influences the polaron distribution.

In fact, $R_{\rm V_{O}-pol}$ decreases with increasing polaron-$V_{\rm O}$ distance, at any reported $c_{V_{\rm O}}$, indicating the overall tendency of polarons to occupy the Ti$^A_{S1}$ sites nearest to the $V_{\rm O}$, which is consistent with recent experimental observation~\cite{Yim2016a}.
This polaron-${V_{\rm O}}$ attraction counteracts the polaron-polaron repulsion and facilitates the short-range arrangement between polarons in different rows, evidenced by the large peaks in $R_{\rm pol-pol}$ at 6.6~\AA\ and 7.0~\AA, as well as in the same row, manifested by the peak at 6~\AA\ (revealing polarons at next-nearest neighbor Ti$^A_{S1}$ sites) [see empty and filled peaks in Fig.~\ref{fig:corr}(a-c)].
In the $S0$ layer, the $V_{\rm O}$-polaron correlation function shows that the preferred site is the Ti$^A_{S0}$ site next nearest neighbor to $V_{\rm O}$, at 4.7~\AA. 
Finally, we note that even though polaron-polaron repulsion hinders the formation of a Ti$^A_{S0}$ polaron directly above a Ti$^A_{S1}$, the presence of an oxygen vacancy mitigates the polaron-polaron repulsive interaction, and formation of such $S0$-$S1$ polaron-complex sporadically occurs (not shown).

At this point it is interesting to compare our data with available experimental data, in particular to acquire some insights on the possible existence of surface $S0$ polarons.
Surface sensitive experimental probes like scanning tunneling microscopy (STM) and scanning tunneling spectroscopy (STS) provide clear evidence of the formation of $S1$ polarons, but do not report any direct indication of $S0$ polaronlike in-gap states, at least at low temperature~\cite{Reticcioli2017c}. 
Similarly to STM/STS, resonant photoelectron diffraction performed at room temperature supports the formation of sub-surface $S1$ polarons, but some diffuse signals coming from surface Ti atoms are detected that were tentatively assigned to $S0$ polarons~\cite{Kruger2008}.
Therefore, one might argue that $S0$ polarons might be activated by temperature. 
As already mentioned, our FPMD are done at 700~K, a choice motivated by the need of increasing the statistics by accelerating the polaron hopping.
As a coupled effect, high-temperature enables $S1$ polarons to overcome the energetic barrier and jump into the $S0$ layer. 
At low $T$ the appearance of $S0$ polarons in FPMD would be a much rarer event, which would require very long time-intervals to be observed.

\begin{figure}[t]
        \includegraphics[width=0.9\columnwidth,clip=true]{\dirimg 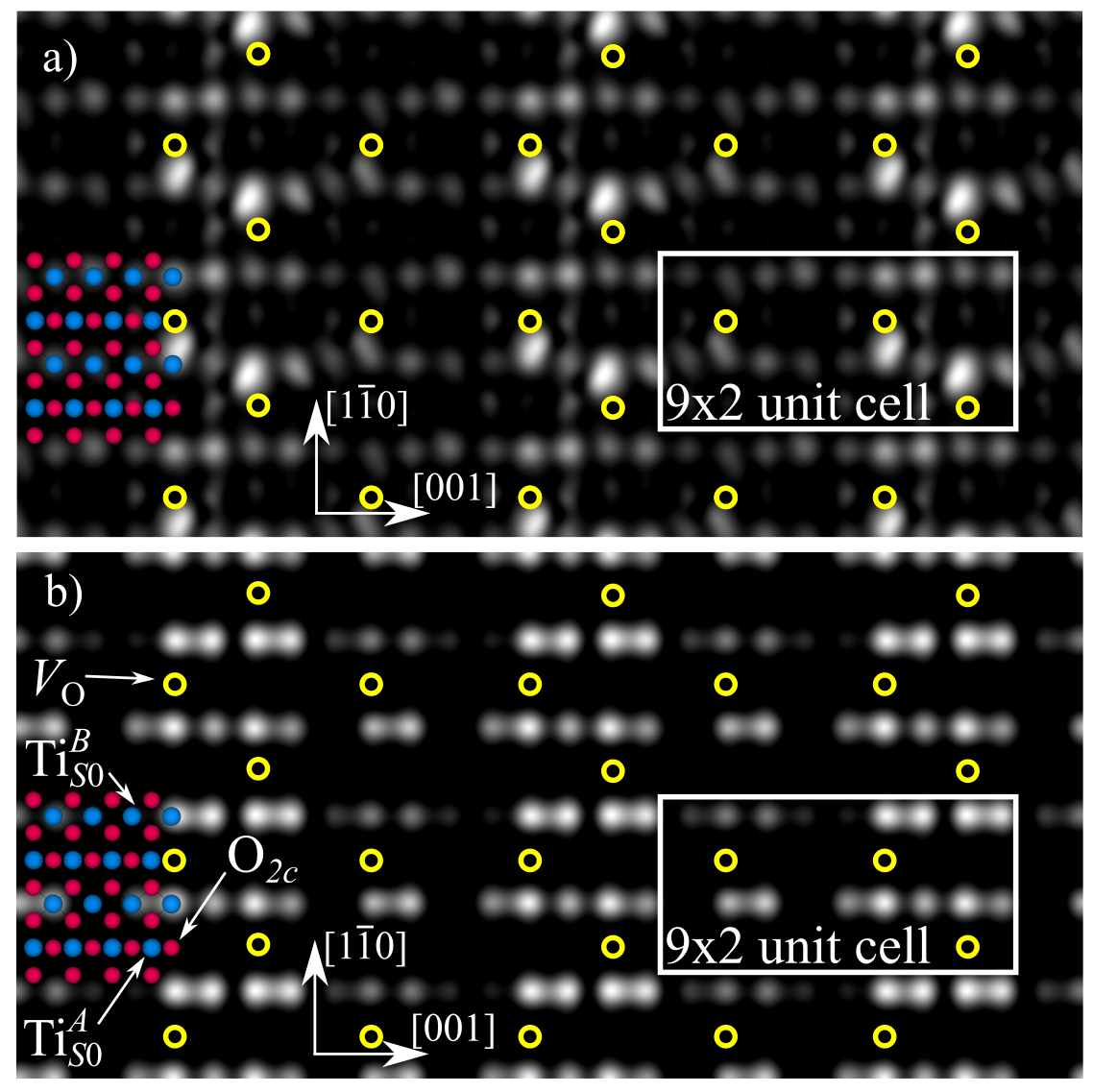}
        \caption{Simulated STM. Post-FPMD DFT+$U$ results for inequivalent polaronic configurations, for a system with $c_{\rm V_O}=16.7\%$. The time-averaged simulated STM images of the in-gap states for the whole FPMD run (a) and for configurations (1168 time steps) with polarons in $S1$ only (b) are shown separately. Empty circles mark the positions of the oxygen vacancies. Ball models representing the $S0$ surface layer are also shown.
}
\label{fig:stm}
\end{figure}

The presence of $S0$ polarons and their influence on experimentally measurable quantities is described well by the simulated STM images shown in Fig.~\ref{fig:stm}. 
Surface $S0$ polarons are prominent in Fig.~\ref{fig:stm}(a), obtained as a time average of the charge of the polaronic states at each FPMD step at the representative concentration $c_{V_{\rm O}}=16.7\%$.
At Ti$^A_{S0}$ sites nearest neighbor to $V_{\rm O}$ (indicated with circles), $S0$ polarons signals appear as particularly bright and diffuse spots.
The less intense and generally circular spots arise from Ti$^A_{S1}$ polarons.
In experimental STM images\cite{Setvin2014, Reticcioli2017c}, it is well-established that Ti$^A_{S1}$ polarons exhibit a dimerlike shape, in apparent disagreement with our simulations derived from high $T$ FPMD. 
To recover the dimerlike feature peculiar of low temperature experiments, we constructed an STM image by averaging the polaronic states over all FPMD-derived configurations owing $S1$ polarons only.
The resulting image, shown in Fig.~\ref{fig:stm}(b), is in excellent agreement with experiment and at the same time satisfy the energetic requirements for favorable Ti$^A_{S1}$ polarons (\textit{i.e.} proximity to the vacancy, and $3\times$1 pattern that maximizes the polaron-polaron distance).

We conclude this part by discussing briefly the results related to the (1$\times$2) reconstructed phase that presents polaron distributions similar to those obtained for the fictitious unreconstructed phase at highly reduced conditions (see Fig.~\ref{fig:histo}):
Charge trapping takes place predominantly in $S0$ rather than $S1$.
Moreover, the Ti atoms in the reconstructed Ti$_2$O$_3$ layer (Ti$_{\rm rec}$) are found to host on average only 1 of the available 18 excess electrons per time step, due to the energetic instability of polarons trapped at the reconstructed sites~\cite{Reticcioli2017c}.
However, the specific site-resolved polaron distribution is different as evidenced in Fig.~\ref{fig:corr}(g,h) where we compare the polaron-polaron correlation obtained for the reconstructed Ti$_2$O$_3$ phase with the corresponding ($c_{V_{\rm O}}=50\%$) unreconstructed one (\textit{i.e.} these two slabs have the same amount of excess electrons).
In the (1$\times$2) phase, excess electrons prefer to be trapped at the Ti$^{\alpha}_{S0}$ sites underneath the reconstructed rows.
Interestingly, at variance with Ti$^A_{S0}$ polarons, excess electrons at Ti$^{\alpha}_{S0}$ sites are even able to occupy adjacent sites.
The reconstruction does not alter the density of polarons in the adjacent Ti$^A_{S0}$ row (1$\times$1 terrace), which on average hosts 3 polarons, as in the (1$\times$1) phase.
The 3$\times$1 pattern is preserved along the Ti$^{\beta}_{S0}$ rows, which correspond to the Ti$^{A}_{S1}$ sites in the (1$\times$1) phase (the Ti$^{\beta}_{S0}$ sites are structurally more similar to the Ti$^{A}_{S1}$ than to the Ti$^{B}_{S0}$ sites).

After having presented the main outcome of the FPMD-based polaron-statistics, in the following section we complement our analysis with the results of the static-model calculations and will provide a global picture on the underlying physics governing the formation and dynamics of polarons in TiO$_2$(110).

\section{Analysis and discussion}
\label{discussion}

As described in the computational section, in order to acquire specific information on the role played by the interaction channels determining the polaron characteristics of TiO$_2$, we disentangled the various interactions and considered the individual contributions by constructing suitable models with well-defined polaron patterns.
The results on the properties of isolated polarons are discussed in Sec.~\ref{results-z}, whereas 
polaron-polaron and polaron-vacancy complexes are discussed in Sec.~\ref{results-polpol} and Sec.~\ref{results-polVO}), respectively. 
Finally, in Sec.~\ref{results-combined} we discuss the combined effects of polaron-vacancy and polaron-polaron interaction in the neutral slab with one ${V_{\rm O}}$ and two polarons.
For the sake of clarity and to allow for a more transparent interpretation of the result, the whole section refers to the (1$\times$1) unreconstructed phase only, at polaron and ${V_{\rm O}}$ concentration within the experimentally measured regimes (0-16.7\%).

\subsection{Isolated polarons.}
\label{results-z}

\begin{table}[h]
\caption{Orbital character (in percentage) of the Ti$^A$ polarons in the $S1$, $S2$ and $S3$ layers; the column `non-local' refers to the amount of charge (in percentage) non localized at the Ti site hosting the polaron. The average bondlength distortion (see Eq.~\ref{eq:distortions}) is also indicated (in \AA\ per atom). Data obtained for independent polarons in the 5$\times$2 slab.
}
\centering
\renewcommand\arraystretch{1.2}
\begin{tabular}{ c || c c c c | c || c } \hline \hline
Polaron & $d_{z^2}$ & $d_{xz}$ & $d_{yz}$ & $d_{x^2-y^2}$ & non-local & $D$ \\ \hline
Ti$^A_{S0}$ & - & 22 & 48 & - & 30 & 0.10 \\
Ti$^A_{S1}$ & 57 & - & - & 14 & 28 & 0.07 \\
Ti$^A_{S2}$ & - & - & - & 71 & 29 & 0.07 \\ \hline \hline
\end{tabular}
\label{table:character}
\end{table}

\begin{figure}[t]
        \includegraphics[width=0.9\columnwidth,clip=true]{\dirimg 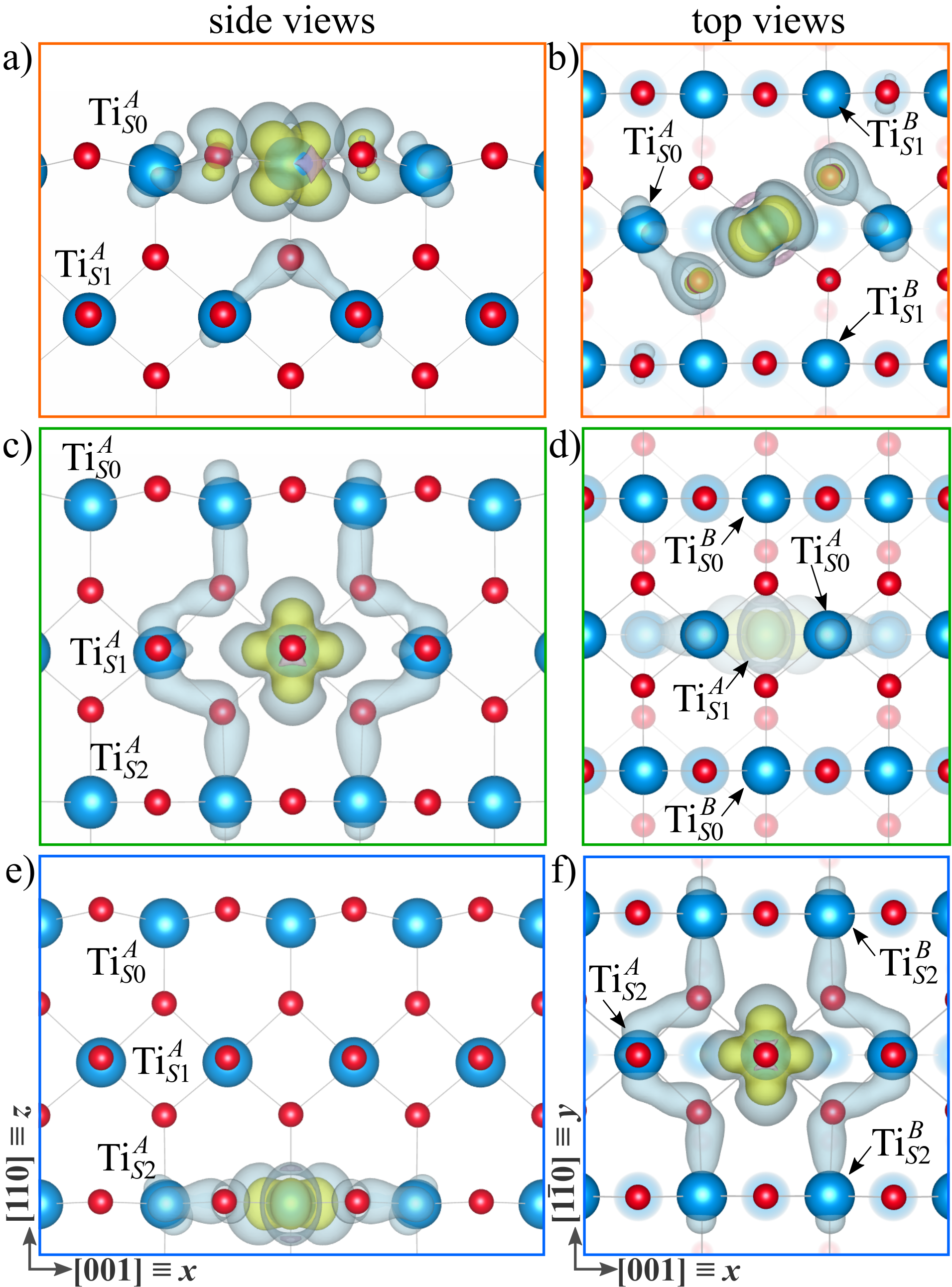}
\caption{Polaron charge density. The side and top views of the Ti$^A_{S0}$  polaron (a,b), Ti$^A_{S1}$  polaron (c,d) and Ti$^A_{S2}$  polaron (e,f) are shown.
The inner and outer isosurfaces represent different levels of the charge density of the polaronic states.
Faded spheres represent deeper atoms in top-view images ($S0$ and $S1$ atoms not shown in panel f).
}
\label{fig:zparchg}
\end{figure}

\begin{figure}[t]
        \includegraphics[width=0.9\columnwidth,clip=true]{\dirimg 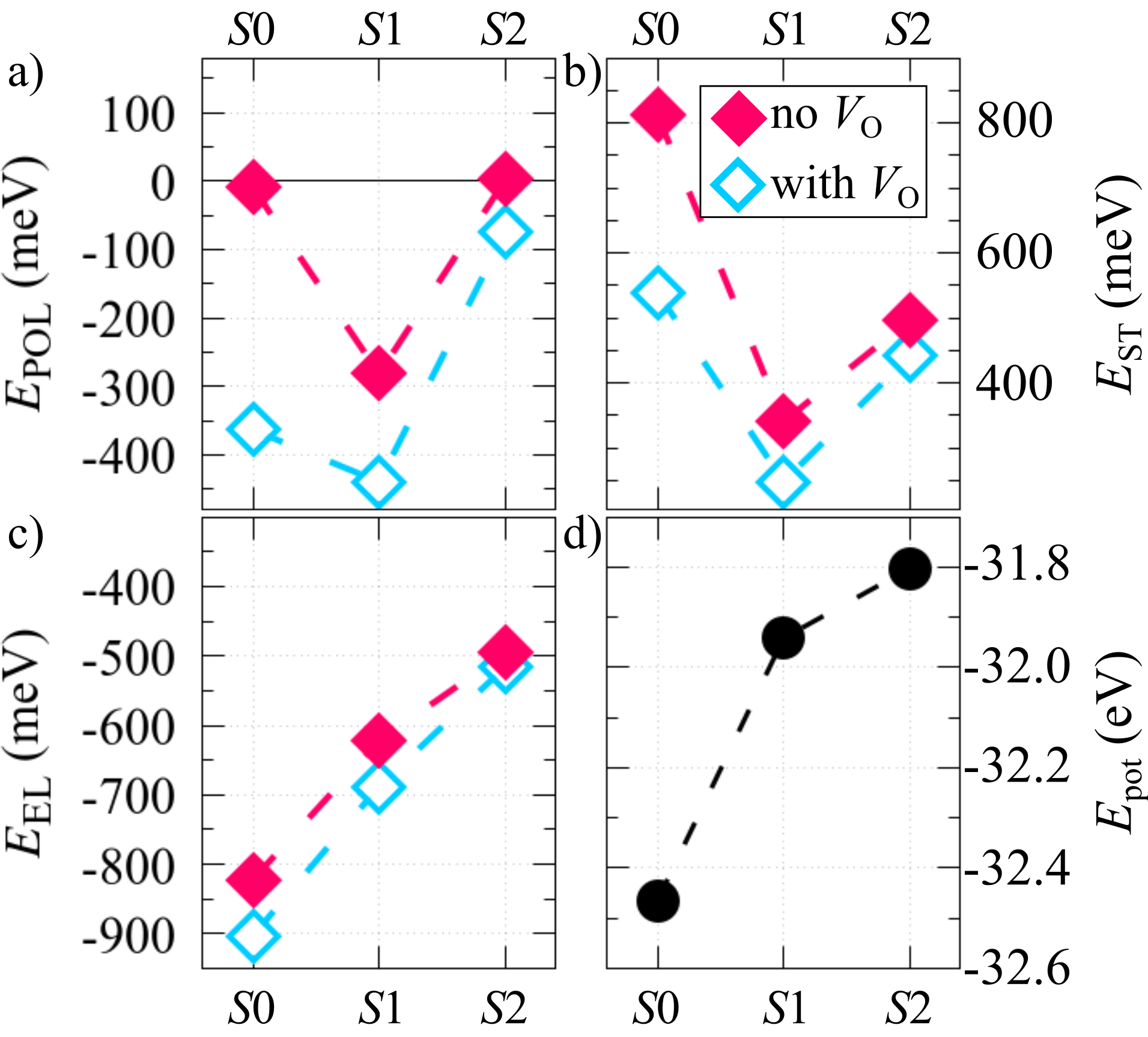}
\caption{Polaron formation energy $E_{\rm POL}$ (a), strain energy $E_{\rm ST}$ (b), and electronic energy $E_{\rm EL}$ (c) of a polaron localized at Ti$^A$ sites at various depths (from $S$0 to $S$2).
Results of stand-alone DFT+$U$ calculations on a 8-layer deep, 3$\times$2-large slab with one excess electron considering both the cases of one and no $V_{\rm O}$ in the cell.
Ti$^A$ sites closest to the $V_{\rm O}$ at the $S1$ and $S2$ layers were considered, while the next-nearest neighbor to the vacancy at $S0$ is shown.
The electrostatic potential energy for the electrons $E_{\rm pot}$ (d) was obtained on a neutral, pristine, 3$\times$2-large, 8-layer deep slab.
}
\label{fig:zenergy}
\end{figure}

The localization of polarons in the surface layers is a key feature of TiO$_2$ rutile but a systematic characterization of the intrinsic properties of isolated polarons is still lacking.
Figure~\ref{fig:zparchg} shows the charge densities for isolated polarons (\textit{i.e.} negligible interactions with $V_{\rm O}$ and other polarons) at various Ti sites (Ti$^A_{S0}$, Ti$^A_{S1}$ and Ti$^A_{S2}$), which are distinct in terms of orbital topology, degree of localization and associated local structural distortions.
The corresponding orbital-projected analysis is reported in Table~\ref{table:character}.
The symmetry of the $d$ orbitals is defined in terms of the $x$, $y$ and $z$ directions which correspond to  [001], [1$\bar{1}$0] and [110], respectively.
To measure the degree of local structural distortions we computed the average bondlength distortion $D$ for the oxygen atoms O$_i$ coordinated to the polaronic Ti site:
\begin{equation}
 D = \frac{1}{N_{\rm O}} \sum_{i=1,N_{\rm O}} |\Delta{{\rm O}_i}|
\label{eq:distortions}
\end{equation}
where $N_{\rm O}$ is the number of O atoms (5 for Ti$^A_{S0}$ polarons, and 6 for Ti$^A_{S1}$ and Ti$^A_{S2}$), and $\Delta{{\rm O}_i} = \delta_{{\rm O}_i}^{loc} - \delta_{{\rm O}_i}^{del}$ is the distortion of the bondlength $\delta$ at each atomic site ${\rm O}_i$ between the localized (polaronic) and delocalized solution.

\textit{Ti$^A_{S0}$ polaron}.
The Ti$^A_{S0}$ polaron is characterized by a predominant $d_{yz}$ orbital character, mixed with a smaller $d_{xz}$ contribution, well recognizable from the polaronic isosurfaces displayed in Fig.~\ref{fig:zparchg}(a,b).
Remarkably, only two thirds of the polaronic charge density is localized at the Ti site.
The remaining 30\% spreads away from the central Ti, and hybridizes asymmetrically with the in-plane oxygen atoms along the $xy$ directions, the two nearest-neighbor Ti atoms along [001], and atoms below the $S0$ layer.
Polaron formation in Ti$^A_{S0}$ induces large structural distortions, mostly localized around the polaron site, quantified by an average bondlength distortion $D$ of $0.10$~\AA\ per O atom. 
The Ti site hosting the polaron relaxes outwards along [110] by 0.14~\AA.
The in-plane nearest-neighbor Ti-O bondlengths involving the O atoms hybridized with the polarons increase by 0.05~\AA~with respect to the positions in the non-polaronic cell, while the remaining two in-plane O are pushed away by 0.11~\AA.
The distortions involve also the surrounding nearest-neighbor Ti atoms, which move outwards the polaronic site by approximately 0.02~\AA. 

\textit{Ti$^A_{S1}$ polaron}.
The Ti$^A_{S1}$ polaron is instead characterized by a dominant $d_{z^2}$ symmetry together with a smaller $d_{x^2-y^2}$ contribution~\cite{Shibuya2012} [Fig.~\ref{fig:zparchg}(c,d)].
One fourth of the polaronic charge density spreads away from the hosting Ti atom:
The hybridization with two Ti atoms in $S0$ determines the dimer-like signal in STM images (see Fig.~\ref{fig:stm}), while the hybridization with the two nearest-neighbor Ti atoms along [001] affects the polaron-polaron interaction (as discussed in Sect.~\ref{results-polpol} and~\ref{results-combined}) and stabilizes the 3$\times$1 pattern at high $c_{V_{\rm O}}$ (discussed in Sect.~\ref{results}).
The Ti$^A_{S1}$ polaron is coupled with small lattice distortions ($D=0.07$~\AA).
The polaronic site moves towards the surface by only 0.03~\AA, the Ti atoms in the same [001] row move inwards by 0.03~\AA, whereas the six octahedrally coordinated O atoms relax outwards by about 0.04-0.08\AA, resulting in rather different in-plane and out-of-plane Ti-O bondlengths.

\textit{Ti$^A_{S2}$ polaron}.
Finally, in the $S2$ layer [Fig.~\ref{fig:zparchg}(e,f)], the Ti$^A_{S2}$ polaron shows a clear $d_{x^2-y^2}$ symmetry.
Due to the 90 degree rotation of the coordination octahedron at the Ti$^A_{S2}$ site with respect to the Ti$^A_{S1}$ site, the polaronic cloud resembles the symmetry of the Ti$^A_{S1}$ but extended in a plane parallel to the (110) surface rather than the (1$\bar{1}$0) plane.
Distortions for the Ti$^A_{S1}$ polaron are small ($D=0.07$~\AA):
The polaron site is very close to the original non-polaronic Ti position, the closest Ti sites move towards the polaronic site by 0.01~\AA, whereas the Ti-O bondlengths undergoes changes of about 0.02-0.09~\AA. 

We also explored the formation of polarons at $B$ sites in $S1$ and $S2$ (not shown), and found that the corresponding polaronic charge have orbital symmetries very similar to those forming at the $A$ sites:
Ti$^B_{S1}$ and Ti$^B_{S2}$ polarons exhibit a $d_{x^2-y^2}$ and $d_{z^2}$ symmetry, respectively, consistently with the orientation of the coordination octahedron of the hosting site.

Summing up, different types of polarons can be formed in TiO$_2$(110) that carry specific orbital symmetries and are coupled with a different degree of lattice distortion.
In addition to the Ti$^A_{S0}$ polaron with $d_{xz}-d_{yz}$ orbitals, we can identify another type of small polarons at $S1$ and deeper layers, with $d_{z^2}$ and $d_{x^2-y^2}$ orbitals, alternately along [110] and [1$\bar{1}$0] as a result of the orientation of the local environment (\textit{i.e.} the polaron occupies a $t_{2g}$ orbital).
The inner charge of $S1$ [Fig.~\ref{fig:zparchg}(c,d)] and $S2$ [Fig.~\ref{fig:zparchg}(e,f)] polarons shows two lobes extending along [001], towards the two nearest-neighbor Ti$^{4+}$ atoms, with consequent decrease of the Ti-Ti bondlength.
In the $S0$ layer, the broken bonds at the surface make the Ti$^A_{S0}$ atoms more negatively charged.
As a result, the inner lobes of $S0$ polarons do not extend along [001], and the two neighbor Ti$^A_{S0}$ sites move away from the polaronic Ti site [Fig.~\ref{fig:zparchg}(a,b)].

Clearly, these different types of polaron are not equally stable and favored. 
The hosting site has a strong influence on the polaron energies.
This is shown in  Fig.~\ref{fig:zenergy}, where we plot the polaron formation ($E_{\rm POL}$), structural ($E_{\rm ST}$) and electronic ($E_{\rm EL}$) energies as well as the electrostatic potential energy $E_{\rm pot}$, for one excess electron in the 3$\times$2, 8-layer-deep slab with and without $V_{\rm O}$ (qualitatively similar behaviors are obtained at lower polaronic concentration, \textit{i.e.} one excess electron in a 5$\times$2 cell, not shown).
The main result is that the $S1$ site is the most favorable one for charge trapping, with some differences due to the presence of a $V_{\rm O}$, and this can be rationalized by inspecting the individual energy contributions, as elaborated below.

In absence of oxygen vacancies, polaron formation is largely favored at $S1$ as compared to the other layers by about $300$~meV [see $E_{\rm POL}$ curve in Fig.~\ref{fig:zenergy}(a)].
When polarons are trapped in $S0$ or $S2$ the energy of the localized polaronic solution $E^{\rm loc}_{\rm relax}$ is almost identical to $E^{\rm deloc}_{\rm relax}$ (the energy of the system with the excess electron delocalized in the conduction band), resulting in essentially no energy gain, \textit{i.e.} $E_{\rm POL}~\approx~0$.
Despite the very large energy gain provided by $E_{\rm EL}$, the formation of an $S0$ polaron is contrasted by a large structural cost $E_{\rm ST}$ [Fig.~\ref{fig:zenergy}(b,c)].
This is due to the large lattice distortions at the undercoordinated atoms around the Ti$^A_{S0}$ polaron (see Table~\ref{table:character}) and to the reduced electron screening at the surface, which leads to an unfavorable energy $E^{\rm deloc}_{\rm constr}$ for the delocalized solution constrained in the distorted structure.
In the $S2$ layer (and deeper layers, not shown) polaron formation is unfavorable due to a reduction of $E_{\rm EL}$ and a still larger structural cost owing to the increased rigidity of the deep layers.
Conversely, charge trapping in $S1$ is preferable due to the relatively small structural cost compared to the the electronic energy gain $E_{\rm EL}$.
Remarkably, $E_{\rm EL}$ follows the trend of the electrostatic potential energy $E_{\rm pot}$ [calculated for the neutral slab, see Fig.~\ref{fig:zenergy}(d)], which is very negative at $S0$ sites (\textit{i.e.} more adapt to attract excess negative charge), and gradually increases at deeper layers (finally saturating at $S2$).

The presence of an oxygen vacancy at the surface partially changes this picture.
Polaron formation in $S1$ is further stabilized ($E_{\rm POL}~\approx~-450$~meV), and the reduced structural cost to distort the lattice at $S0$ (from 800~meV to 500~meV) results in a rather large $E_{\rm POL}$, making polaron formation in $S0$ a more favorable process as compared to the vacancy-free situation. 
This strong reduction of the structural costs does not involve sub-surface layers, where $E_{\rm ST}$ changes only slightly (in the range of 20~meV), which is however sufficient to maintain  $S1$ the most favorable site for polaron trapping.

Oxygen vacancies affect also the energy barrier for a polaron to hop between two sites at different layers.
In the case of a pristine surface, the energy barriers for polaron-hopping from $S1$ to $S0$ and from $S1$ to $S2$  are relatively large, about 450~meV.
The inclusion of a $V_{\rm O}$ decreases significantly the $S1$-to-$S0$ barrier, to $\approx$ 300~meV (a value similar to the one found from the post-FPMD model, as discussed in Sec.~\ref{results}), whereas the barrier for $S1$-to-$S2$ polaron hopping increases, 500~meV.
These results indicate that $V_{\rm O}$ plays the role of an attracting center for polaron.
This issue will be inspected in more detailed in Sec.~\ref{results-polVO}.

To conclude this part, we highlight that the above analysis is consistent and explains (in terms of energy balance) the trend of the distribution of polarons among the various layers reported in Fig.~\ref{fig:histo} and already discussed in Sec.~\ref{results}.
At low oxygen vacancy concentrations ($c_{V_{\rm O}} \leq 16.7\%$) polarons forms preferentially at the $S1$ layer, with respect to $S0$ and $S2$.
At larger $c_{V_{\rm O}}$ they are gradually more attracted by $S0$ sites, and this is also favored by the relatively small structural costs to further distort the lattice at the surface.
We recall that in our FPMD runs we used a 5-layer deep cell (with the 2 layers at the bottom kept fixed).
This setup weakens $E_{\rm POL}$ and restrains the formation of polarons in deep layers but is a physically valid choice due to the inconvenience to form polarons at layers deeper than $S1$, as demonstrated by the results based on the static-model approach.

\begin{figure}[h!]
    \begin{center}
        \includegraphics[width=0.9\columnwidth,clip=true]{\dirimg 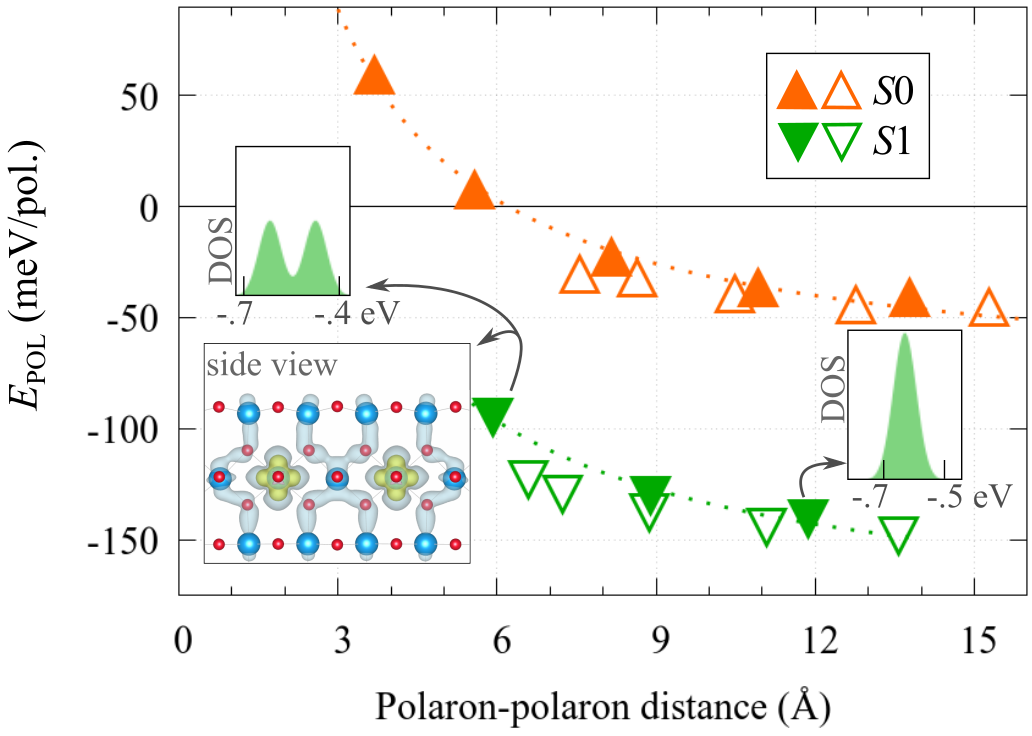}
    \end{center}
\caption{Polaron-polaron interaction. $E_{\rm POL}$ (in meV per polaron) as a function of the distance between two polarons in the pristine 9$\times$2 slab ($-2$ charged slab). One polaron is fixed at a Ti$^A_{S1}$ site, with the other one sitting in Ti$^A$ sites in $S$0 (up-pointing triangles) and $S$1 (down-pointing triangles).
Filled symbols refer to sites lying in the same ($1\bar{1}0$) layer (perpendicular to the surface layers) as the reference polaron fixed at Ti$^A_{S1}$, whereas empty symbols indicate sites in the nearest-neighbor ($1\bar{1}0$) layer formed by Ti$^A$ atoms.
The insets show the in-gap DOS polaronic peaks for two particular configurations (energy values in eV units, with respect to the bottom of the conduction band).
The spatial distribution of the polaronic charge is also shown for polarons located at two next-nearest neighbor Ti$^A_{S1}$ sites (two lattice constants apart, $\approx$~6\AA).
}
\label{fig:Epolpol}
\end{figure}

\subsection{Polaron-polaron interaction.}
\label{results-polpol}

This section focuses on the quantitative analysis of the polaron-polaron interaction and its effect on the overall energetics.
Figure~\ref{fig:Epolpol} collects the results obtained by the static-model approach for a pristine 9$\times$2  slab containing two excess electrons.
One polaron is kept fixed at a Ti$^A_{S1}$ site, while the second one is treated as test-polaron systematically localized at different Ti$^A$ sites in $S1$ and $S0$.
Consistent with previous studies~\cite{Deskins2011} and in agreement with the results obtained with a deeper slab and presented in Sec.~\ref{results-z}, the total energy of the system is lower when both polarons are localized in ${S1}$ (see down-pointing triangles).

The polaron energy $E_{\rm POL}$ depends strongly on the distance between the test-polaron and reference polaron fixed at Ti$^A_{S1}$, regardless on the specific layer. In both $S1$ and $S0$, $E_{\rm POL}$ decreases with increasing polaron-polaron distance.
First, we note that due to the strong electronic repulsion, we could not obtain a configuration with the two polarons localized at adjacent Ti$^A_{S1}$ sites (2.97~\AA) along [001] in a pristine cell, unless forcing large local lattice distortions by using a larger $U$ of 5~eV.
The spatially smallest polaron-polaron complex is the one with the test-polaron located 2 Ti sites apart from the reference polaron, corresponding to a polaron-polaron distance of 5.93~\AA.
The energy gain to separate the polarons lying along the same [001] row (filled $S1$ triangles) at a distance of four lattice constants (11.87~\AA) is quite large, 46~meV, clearly indicating that polarons prefer to be spatially separated.
The trend of $E_{\rm POL}$ as a function of the polaron-polaron distance is useful to determine the minimal setup to adopt in simulations aiming to describe isolated polarons:
By using less than three lattice sites along [001], the overlap of the polaronic charge-density clouds is large, and the spurious interactions between a polaron and its periodical image are not negligible~\cite{Calzado2008}.

If the two polarons are localized in two adjacent Ti$^A_{S1}$ [001] rows (empty $S1$ triangles), variations on $E_{\rm POL}$ are rather small (25~meV), and mostly attributable to the structural $E_{\rm ST}$ contribution ($E_{\rm EL}$ varies only slightly, not shown, since screening effects weaken the polaron-polaron Coulomb repulsion at large distance).

Importantly, the increasing stability with increasing distance is reflected in the location and topology of the polaronic level within the gap region.
At larger separation the polarons do not interact among each other and form independent polaron peaks, which are degenerate in energy.
Conversely, spatially confined polaronic pairs split this degeneracy and give rise to a double-peak structure due to the enhanced  polaron-polaron interaction.
This issue will be discussed in more detail in Sec.~\ref{results-combined}.

An analogous trend is observed when the two polaron are localized in different layers, at  Ti$^A_{S1}$ and Ti$^A_{S0}$ sites (up-pointing triangles).
In this case we could form a spatially confined polaron-polaron complex within a distance of $\approx$~4~\AA\ (one polaron above the other), but the resulting $E_{\rm POL}$ is positive, indicating an intrinsic instability of this solution. 
This is due to the repulsive interaction between the $S0$ polaron with the electronic cloud of the $S1$ polaron, which spreads towards the surface Ti$^A_{S0}$ sites right above and leads to a reduction of the electronic energy gain $E_{\rm EL}$.
This is not compensated by a corresponding gain in the structural energy, even though smaller lattice distortions are required to accommodate spatially confined polaron-pairs. 
Maximizing the distance between the two polarons, 15~\AA, results in an overall energy gain of more than 100~meV.
Similarly to the $S1$ case, the variation of $E_{\rm POL}$ is small (15~meV) if the two polarons are located in more distant [001] rows (empty $S0$ triangles).

\subsection{Polaron-$V_{\rm O}$ interaction.}
\label{results-polVO}

\begin{figure}[t]
    \begin{center}
        \includegraphics[width=0.9\columnwidth,clip=true]{\dirimg 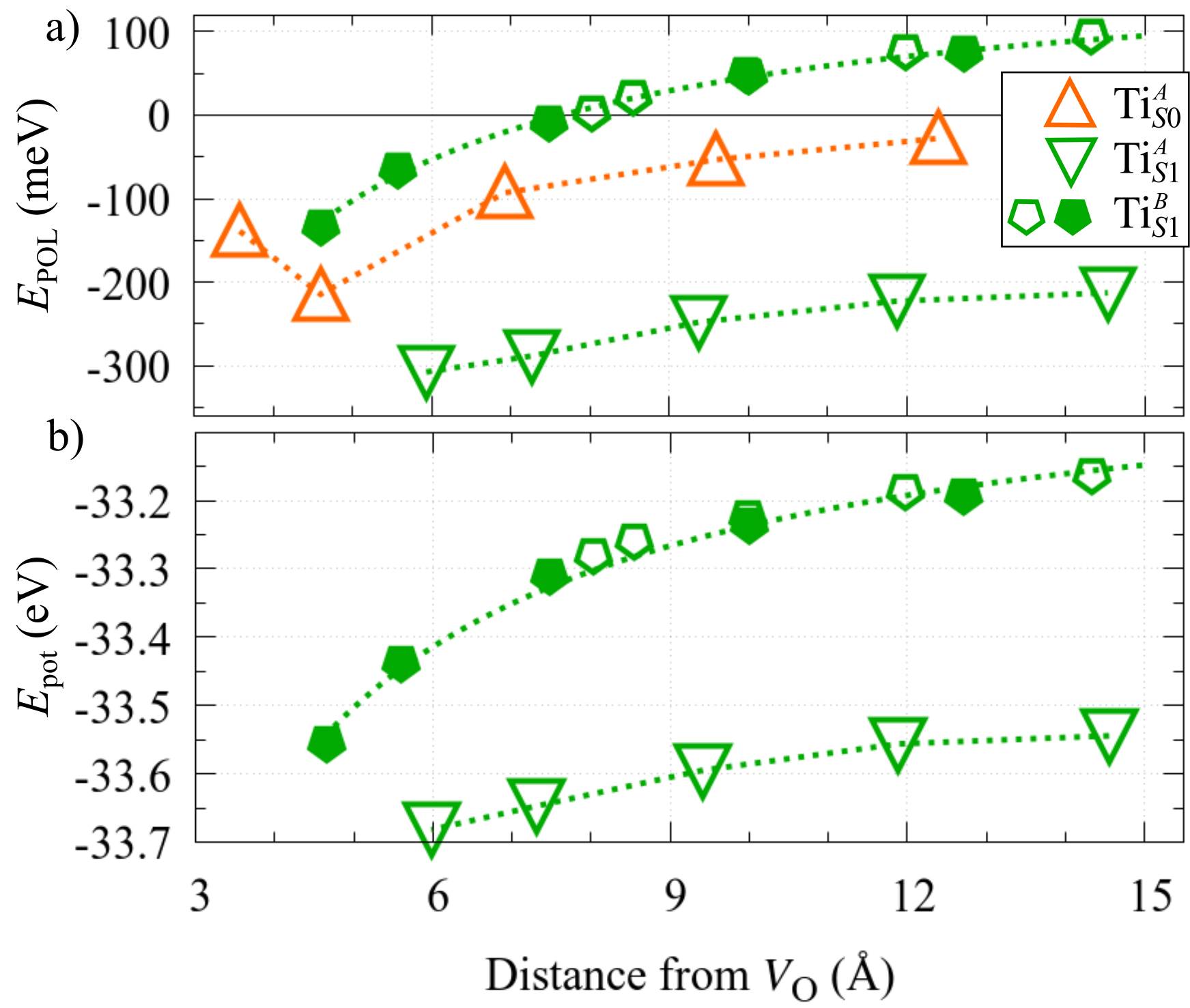}
    \end{center}
\caption{Polaron-vacancy interaction. Panel (a): $E_{\rm POL}$ as a function of the distance between the oxygen vacancy and the polaron, included in the 9$\times$2 slab (\textit{i.e.} $c_{\rm V_O}=5.6\%$, $+1$ charged system). The polaron explores the Ti$^A_{S0}$, Ti$^A_{S1}$ and Ti$^B_{S1}$ sites. Defect states at Ti$^B_{S0}$ sites adjacent to the $V_{\rm O}$ (not reported) show a large positive $E_{\rm POL}$ ($\simeq 300$~meV). Panel (b): $E_{\rm pot}$ on Ti$^A_{S1}$ and Ti$^B_{S1}$ sites as a function of a distance from the oxygen vacancy, in the $+2$ charged slab (one $V_{\rm O}$ and no excess electrons).
In both panels, filled and empty symbols represent polaron positions with the same or different [$1 \bar 10$] coordinate with respect to $V_{\rm O}$, respectively.
}
\label{fig:EpolVO}
\end{figure}

As mentioned above, the removal of an oxygen O$_{2c}$ atom from the surface creates a positively charged vacancy that it is expected to behave as an attractive center for the negatively charged polarons.
Figure~\ref{fig:EpolVO} offers an overview on the effects of the $V_{\rm O}$ on an individual polaron. 
To model this situation we consider the reduced 9$\times$2 slab containing one single $V_{\rm O}$ and an extra hole that neutralizes one of the two excess electrons provided by the $V_{\rm O}$.
These data confirm that polaron formation is energetically favored in proximity of the vacancy~\cite{Moses2016,Shibuya2014} for essentially all considered sites [Fig.~\ref{fig:EpolVO}(a)].
Forming a  polaron at the Ti$^A_{S1}$ site closest to the $V_{\rm O}$ is 95~meV more favorable than for large (14.54~\AA) polaron-$V_{\rm O}$ distances.

The polaron-$V_{\rm O}$ attractive interaction influences particularly the propensity of Ti$^B_{S1}$ sites to host polarons. 
As already mentioned, in the pristine cell the formation of polarons in $B$ sites is unstable ($E_{\rm POL} > 0$).
The presence of a surface vacancy reduces the strain cost and increases the electronic gain (not shown) associated with the Ti$^B_{S1}$ polarons, making polaron formation at $B$ sites possible (negative $E_{\rm POL}$) for polaron-$V_{\rm O}$ distances smaller than approximatively 8~\AA.

The trend of $E_{\rm POL}$ correlates well with the electrostatic potential.
This is shown in Fig.~\ref{fig:EpolVO}(b) by comparing the electrostatic potential energy $E_{\rm pot}$ for Ti $A$ and $B$ sites in S1.
At long distance from the $V_{\rm O}$, $E_{\rm pot}$ is largely more negative at Ti$^A_{S1}$ sites (more suitable for polaron formation) as compared to Ti$^B_{S1}$ sites, due to a different degree of local-structure distortions induced by the broken symmetry at the surface.
By decreasing the distance from the $V_{\rm O}$, $E_{\rm pot}$ decreases quickly at Ti$^B_{S1}$ sites, and stabilizes the polaron formation (negative $E_{\rm POL}$).
Thus, Ti$^A_{S1}$ and Ti$^B_{S1}$ sites, despite being similar in terms of the local structural coordination and possibly geared to host polarons with similar orbital symmetry ($d_{z^2}$ and $d_{x^2-y^2}$, respectively), do have a very different $E_{\rm POL}$ as a consequence of the very different electrostatic potential.

$S0$ sites follow a similar trend of $E_{\rm POL}$, with the exception of the Ti$^A_{S0}$ site nearest neighbor to $V_{\rm O}$ which is energetically less favorable than the next nearest neighbor one~\cite{Deskins2011}, as evidenced by the kink at about 4.5~\AA.
Moreover, the nearest-neighbor Ti$^A_{S0}$ polaron retains a dominant $d_{x^2-y^2}$ orbital character ($52\%$), at variance with all other Ti$^A_{S0}$ polarons that show a $d_{xz}$-$d_{yz}$ symmetry. 

Consistent with the trends of $E_{\rm POL}$, the distinct polaronic state (not shown) lies at different energies in the gap region.
In fact, the polaronic state depends on the type of the hosting site and on the interweaved interaction among polarons as well as on the interaction between polarons and $V_{\rm O}$, which strongly depend on the relative positions.
Therefore, by considering only the Ti$^A_{S1}$ polarons observed by experiments~\cite{Kruger2008,Reticcioli2017c}, our data suggest that modifications on the energy of the polaronic in-gap state should be interpreted as a result of the interaction with other polarons and oxygen vacancies, rather than an intrinsic property of the isolated polaron~\cite{Sezen2015a}.
We will further discuss the DOS structure in the next section.

\subsection{Combined $V_{\rm O}$-polaron and polaron-polaron effects.} 
\label{results-combined}

\begin{figure*}[t]
        \includegraphics[width=1.9\columnwidth,clip=true]{\dirimg 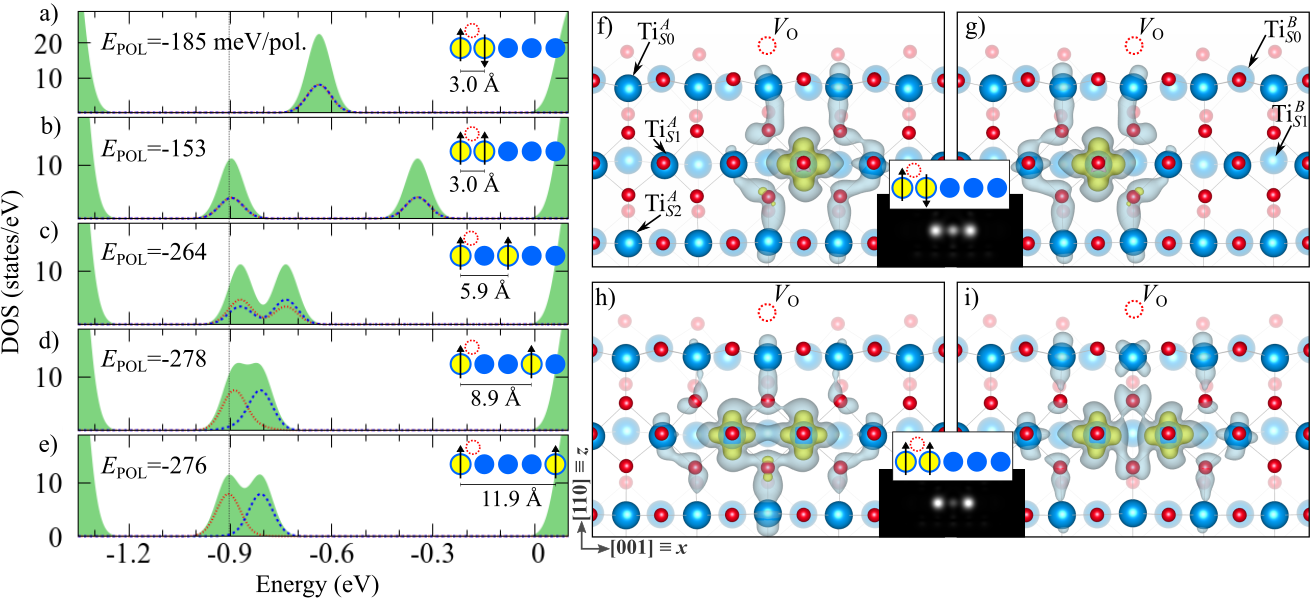}
\caption{$V_{\rm O}$-polaron and polaron-polaron combined effect for polarons in Ti$^A_{S1}$ sites. Panels (a-e): DOS calculated for two polarons at various distances with opposite (a) and parallel (b-e) spins.
The total DOS and the projection on the two polaronic sites are shown with filled and dotted curves, respectively. One polaron is fixed at the Ti$^A_{S1}$ closest to $V_{\rm O}$ while the other one is localized at various sites of the same [001] Ti row.
The respective $E_{\rm POL}$ is indicated on each panel (in meV per polaron).
Panels (f-i): spatial charge density of the polaronic states for two adjacent polarons with opposite (f,g) and parallel (h,i) spins.
Each state is shown separately: Spin up (f) and spin down (g) charge densities for the antiferromagnetic configuration, and the charge densities of the $-0.90$~eV (h) and $-0.35$~eV (i) energy states for the parallel-spin configuration.
The insets show the simulated STM resulting from the in-gap states.
}
\label{fig:polpol-polVO}
\end{figure*}

This final section focuses on the combined polaron-polaron-vacancy effects in the neutral system, containing one $V_{\rm O}$ and two polarons, modeled with  the standard 9$\times$2 slab.

Besides facilitating individual charge trapping at the surface and sub-surface sites, the presence of an oxygen vacancy weakens the strong repulsion between polaron pairs localized at adjacent Ti$^A_{S1}$ sites along [001].
This allows us to study the evolution of the polaronic properties as a function of the polaron-polaron distance from very short to large separation.
The results are collected in Figure~\ref{fig:polpol-polVO}.

Although formation of adjacent polarons is clearly not one of the most favorable configurations, $E_{\rm POL}$ is negative [see Fig.~\ref{fig:polpol-polVO}(a,b)], and the energy gain is influenced by the orientation of the polaronic-spins~\cite{Deskins2011,DiValentin2006,Mackrodt1997}.
A polaron can be described as a localized spin and when the distance between polarons is reduced it is important to take into account the magnetic ordering of the polaron complexes.
We do so by testing two different spin configurations, parallel and antiparallel.
As expected, when the spins are separated by only one lattice constant the antiferromagnetic state (resulting in a zero total magnetization) is the energetically more favorable solution and is characterized by two polaronic peaks at the same energy, $\approx 0.65$~eV below the bottom of the conduction band [two-spin channel DOS in Fig.~\ref{fig:polpol-polVO}(a)].
In contrast, the parallel alignment induces a large splitting between the two polaronic peaks of about 0.6~eV [Fig.~\ref{fig:polpol-polVO}(b)].
The two distinct spin-polarized solutions exhibit different orbital symmetries, as graphically visualized in the spin-dependent isosurfaces shown in Fig.~\ref{fig:polpol-polVO}(f-i).
In the anti-parallel spin case, the orbital occupation of one polaron is not drastically altered by the overlap with the electronic cloud of the other polaron in the neighboring site which belongs to a different spin channel.
Therefore, the two polarons individually retain the same $d_{z^2}$ orbital character, typical of Ti$^A_{S1}$ polarons [Fig.~\ref{fig:polpol-polVO}(f,g)].
Conversely, within the parallel-spin configuration, a hybridization of the polaronic state occurs, which causes a spread of the polaron charge across both adjacent sites:
The polarons are no longer fully localized in one specific Ti site, rather, the polaronic charge is shared between the two adjacent sites in a sort of bonding/anti-bonding configuration 
[see partial DOS in Fig.~\ref{fig:polpol-polVO}(b) and isosurfaces in Fig.~\ref{fig:polpol-polVO}(h,i)].
The spin-integrated STM signals of the adjacent polaron-pairs are qualitatively similar for the two spin alignments [insets in Fig.~\ref{fig:polpol-polVO}(f,g) and Fig.~\ref{fig:polpol-polVO}(h,i)], both very different from a superposition of two double-lobed shape typical of individual polarons [see Fig.\ref{fig:stm}(b)].
They are characterized by a weak spot in the middle between two bright ones, separated by two lattice constants. 
Only spin-dependent STM would be able to experimentally detect the difference between the parallel and antiparallel orderings. 
In fact, the two brightest spots in the antiparallel case come from one specific spin channel (compare the polaron isosurfaces in the top layer in Fig.~\ref{fig:polpol-polVO}(f,g)).    
But, as already highlighted, this is not a likely polaronic configuration in TiO$_2$(110), while it was experimentally observed in materials with a high density of polarons and a similar lattice structure.~\cite{Lakkis1976}

The spin-dependent splitting of the polaronic state vanishes very rapidly with the distance, due to the reduction of the overlap of the polaronic clouds.
For next-nearest-neighbor spins the polaronic energy difference between parallel and antiparallel orderings reduces to less than 1~meV. In the following we therefore only discuss the parallel-spin solution.

By separating the polarons by one additional site along [001] (\textit{i.e.}, two Ti$^A_{S1}$ atoms 5.9~\AA\ apart), the splitting reduces to 0.1~eV and the spin-dependent polaron charge is almost equally distributed among the two Ti sites [see Fig.~\ref{fig:polpol-polVO}(c)].
As compared to the analogous configuration of two polarons located 5.9~\AA\ apart without oxygen vacancy (inset in Fig.\ref{fig:Epolpol}), the energy separation between the two polaron peaks is not largely affected by the introduction of the $V_{\rm O}$, confirming that the nature of the splitting should be attributed to the polaron-polaron hybridization.
However, in the presence of a $V_{\rm O}$, both states are shifted towards deeper energy by 0.3~eV, leading to an enhanced $E_{\rm POL}$ ($-264$~meV rather than $-94$~meV observed in the slab without $V_{\rm O}$, see Fig.\ref{fig:Epolpol}), with a major contribution arising from the electrostatic potential, and only a slight reduction of the structural cost.

By further increasing the distance along [001], the two polarons can be effectively considered as independent from each other.
In fact, the charge overlap is negligible for inter-polaron distances of at least three Ti$^A_{S1}$ sites, with a consequent reduced repulsion between the polarons.
The splitting is strongly reduced, and each polaron is localized around its Ti$^A_{S1}$ trapping center [Fig.~\ref{fig:polpol-polVO}(d,e)], mimicking the situation of antiparallel aligned adjacent polarons [Fig.~\ref{fig:polpol-polVO}(a,f,g)]. 
It is important to note that the residual splitting between the two peaks for a distance of 8.9~\AA\ and 11.9~\AA\ does not originate from hybridization effects, but rather from electrostatic effects due to the fact that polarons are trapped in inequivalent positions with respect to the $V_{\rm O}$.
$E_{\rm POL}$ saturates to a value of about -280~meV, since both attractive ($V_{\rm O}$-polaron) and repulsive (polaron-polaron) interaction decay rapidly with increasing separations.
Consistently, the deepest polaronic peak remains located at about $-0.9$~eV, the typical energy level associated to individual polaronic states in proximity to a $V_{\rm O}$.
As mentioned before, we found that in defect-free samples the polaron is less stable and the characteristic polaronic peak is instead typically located at about 0.6~eV below the bottom of the conduction band.

\section{Summary and Conclusions}
\label{conclusions}

In this work we have analyzed the conditions for the formation of polarons and their dynamics in the reduced rutile TiO$_2$(110) surface by using first principles static and dynamic calculations in the framework of DFT+$U$ and FPMD (at $T=700$~K).
The excess electrons were obtained by removal of oxygen atoms in the surface layer, and a wide range of $V_{\rm O}$ concentrations were considered from $c_{V_{\rm O}}=0$ [pristine (1$\times$1)] to $c_{V_{\rm O}}=50\%$.

During the FPMD runs thermally activated hopping drives the polarons to a wide variety of configurations.
At low $c_{V_{\rm O}}$, polarons reside predominantly at the $S1$ sub-surface layer, which is energetically favored.
At $c_{V_{\rm O}}=16.7\%$, the polarons tend to populate the $S1$ layer with an optimal 3$\times$1 pattern, which avoids the repulsive interaction of negatively charged electronic clouds occurring at short polaron-polaron distance. 
At higher concentrations ($c_{V_{\rm O}}>16.7\%$), the distribution of polarons in the system is too high to preserve the 3$\times$1 pattern in $S1$, and polarons start to form in $S0$ sites.
At higher concentration the high density of polarons in the surface and subsurface layers destabilizes the 1$\times$1 surface and the systems undergoes a (1$\times$2) structural reconstruction, which is able to host a larger number of polarons~\cite{Reticcioli2017c}.

Each inequivalent polaronic configuration observed in the FPMD was further analyzed with DFT+$U$ at $T=0$~K, in order to calculate $E_{\rm POL}$ and understand the energy balance.
To disentangle the different effects contributing to the degree of stability of the various polaron configurations, and to address the role of polaron-polaron and polaron-vacancy interactions, we employed static models, in which we selectively varied the number and and relative distance of polarons and $V_{\rm O}$'s in the (1$\times$1) system.
Moreover, we engineered the position of the polarons, in order to explore a large ensemble of configurations, and compare the resulting polaronic properties.
These results provide a clear and comprehensive picture.

We found that various Ti sites are able to host polarons, which differ by their specific location in the slab and by the different local chemical and structural environment.
This gives rise to different types of polarons with distinct characteristics in terms of orbital symmetry, spatial localization and local structural distortions.
In the (1$\times$1) phase, polaronic Ti$^A_{S0}$ sites are accompanied by larger local distortions of the lattice, with a polaronic cloud exhibiting a $d_{xz}$-$d_{yz}$ orbital.
At deeper layers, $A$ and $B$ Ti sites are able to host $d_{z^2}+d_{x^2-y^2}$ and $d_{x^2-y^2}$ polarons, alternately in the [110] and [1$\bar{1}$0] directions, consistently with the orientation of the coordination octahedra.
The Ti$^A_{S1}$ atoms are the most energetically stable sites for isolated polarons, in agreement with experimental observations. 
The electrostatic potential at Ti$^A_{S1}$ sites is stronger, owing to the local-structure distortions induced by the broken symmetry at the surface, and determines an electron energy gain $E_{\rm EL}$ large enough to overcome the small strain energy cost $E_{\rm ST}$ required to locally distort the lattice and accommodate a polaron.
Conversely, the Ti$^A_{S0}$ sites suffer from a large strain energy cost $E_{\rm ST}$ to distort the lattice in the defect-free surface, while formation of polarons at Ti$^B_{S1}$ sites is hindered by an unfavorable electrostatic potential.
The presence of a vacancy on the surface increases the flexibility of the lattice, thereby lowering the $E_{\rm POL}$ in both surface and sub-surface Ti sites.
In general, we found that $V_{\rm O}$'s act as attractive centers for polarons (a situation also observed in other oxides~\cite{Bondarenko2015}), reduce $E_{\rm ST}$, increase the electronic gain $E_{EL}$ (due to an attractive electrostatic potential), and can influence the orbital symmetry of the neighboring polarons.

The polaron-polaron interaction is clearly repulsive and is particularly effective at small distances.
Polaron pairs at nearest-neighbor Ti$^A_{S1}$ sites along [001] only form in proximity of an oxygen vacancy, whose electrostatic potential mitigates the strong polaron-polaron repulsion enhanced by the overlap of the polaronic charges.
In this configuration, the energy level of the characteristic in-gap polaron peak as wells as its shape depend significantly on the spin alignment of the two polarons.
The antiferromagnetic configuration is energetically more favorable and results in polarons at degenerate energy levels, with electronic clouds resembling the features of isolated Ti$^A_{S1}$ polarons.
Conversely, for ferromagnetically aligned spins, the two polarons are shared equivalently by the two hosting Ti$^A_{S1}$ sites, and the polaronic orbitals undertake a substantial hybridization, with a bonding/anti-bonding splitting of about 0.6~eV.
At larger inter-polarons distances (\textit{e.g.} more than 2 sites apart along the [001] row), the charge overlap becomes negligible: antiferro and ferro solutions are degenerate in energy and the polaron charge remains distinctively localized in one single Ti$^A_{S1}$ sites.

In conclusion, the results presented here offer a valid key to interpret the behavior of small polarons in TiO$_2$ and we believe are representative of the general behavior of polaron in oxides.
In particular, in oxide surfaces the site-specific polaron characteristic will influence the interaction with adsorbates and is expected to play a crucial role in catalysis.
These issues will be discussed in future works. 

From a technical point of view our study confirms the need to control the charge trapping process in simulations rather than relying on the spontaneous electron localization, which, in same cases, could not lead to the global minimum of the system.
Finally, we would like to underline the importance of adopting large supercells in other to either minimize the spurious overlap among polaronic orbitals and avoid rigid constrains on the polaron-induced lattice relaxations.


\begin{acknowledgement}
This work was supported by the Austrian Science Fund (FWF) SFB project VICOM (Grant No. F41),
by the FWF project POLOX (Grant No. I 2460-N36),
by the ERC Advanced Research Grant `OxideSurfaces',
and
by the FWF Wittgenstein-prize (Z250-N16).
The computational results presented have been achieved using the Vienna Scientific Cluster (VSC).
\end{acknowledgement}

\bibliography{bib-2018-04-27}

\providecommand{\latin}[1]{#1}
\providecommand*\mcitethebibliography{\thebibliography}
\csname @ifundefined\endcsname{endmcitethebibliography}
  {\let\endmcitethebibliography\endthebibliography}{}
\begin{mcitethebibliography}{60}
\providecommand*\natexlab[1]{#1}
\providecommand*\mciteSetBstSublistMode[1]{}
\providecommand*\mciteSetBstMaxWidthForm[2]{}
\providecommand*\mciteBstWouldAddEndPuncttrue
  {\def\EndOfBibitem{\unskip.}}
\providecommand*\mciteBstWouldAddEndPunctfalse
  {\let\EndOfBibitem\relax}
\providecommand*\mciteSetBstMidEndSepPunct[3]{}
\providecommand*\mciteSetBstSublistLabelBeginEnd[3]{}
\providecommand*\EndOfBibitem{}
\mciteSetBstSublistMode{f}
\mciteSetBstMaxWidthForm{subitem}{(\alph{mcitesubitemcount})}
\mciteSetBstSublistLabelBeginEnd
  {\mcitemaxwidthsubitemform\space}
  {\relax}
  {\relax}

\bibitem[Stoneham(1989)]{Stoneham1989}
Stoneham,~A.~M. {Small polarons and polaron transitions}. \emph{Journal of the
  Chemical Society, Faraday Transactions 2} \textbf{1989}, \emph{85}, 505\relax
\mciteBstWouldAddEndPuncttrue
\mciteSetBstMidEndSepPunct{\mcitedefaultmidpunct}
{\mcitedefaultendpunct}{\mcitedefaultseppunct}\relax
\EndOfBibitem
\bibitem[Shluger and Stoneham(1993)Shluger, and Stoneham]{Shluger1993}
Shluger,~A.~L.; Stoneham,~A.~M. {Small polarons in real crystals: concepts and
  problems}. \emph{Journal of Physics: Condensed Matter} \textbf{1993},
  \emph{5}, 3049--3086\relax
\mciteBstWouldAddEndPuncttrue
\mciteSetBstMidEndSepPunct{\mcitedefaultmidpunct}
{\mcitedefaultendpunct}{\mcitedefaultseppunct}\relax
\EndOfBibitem
\bibitem[Stoneham \latin{et~al.}(2007)Stoneham, Gavartin, Shluger, Kimmel,
  Ramo, R{\o}nnow, Aeppli, and Renner]{Stoneham2007}
Stoneham,~A.~M.; Gavartin,~J.; Shluger,~A.~L.; Kimmel,~A.~V.; Ramo,~D.~M.;
  R{\o}nnow,~H.~M.; Aeppli,~G.; Renner,~C. {Trapping, self-trapping and the
  polaron family}. \emph{Journal of Physics: Condensed Matter} \textbf{2007},
  \emph{19}, 255208\relax
\mciteBstWouldAddEndPuncttrue
\mciteSetBstMidEndSepPunct{\mcitedefaultmidpunct}
{\mcitedefaultendpunct}{\mcitedefaultseppunct}\relax
\EndOfBibitem
\bibitem[Nagels \latin{et~al.}(1963)Nagels, Denayer, and Devreese]{Nagels1963}
Nagels,~P.; Denayer,~M.; Devreese,~J. {Electrical properties of single crystals
  of uranium dioxide}. \emph{Solid State Communications} \textbf{1963},
  \emph{1}, 35--40\relax
\mciteBstWouldAddEndPuncttrue
\mciteSetBstMidEndSepPunct{\mcitedefaultmidpunct}
{\mcitedefaultendpunct}{\mcitedefaultseppunct}\relax
\EndOfBibitem
\bibitem[Verdi \latin{et~al.}(2017)Verdi, Caruso, and Giustino]{Verdi2017}
Verdi,~C.; Caruso,~F.; Giustino,~F. {Origin of the crossover from polarons to
  Fermi liquids in transition metal oxides}. \emph{Nature Communications}
  \textbf{2017}, \emph{8}, 1--7\relax
\mciteBstWouldAddEndPuncttrue
\mciteSetBstMidEndSepPunct{\mcitedefaultmidpunct}
{\mcitedefaultendpunct}{\mcitedefaultseppunct}\relax
\EndOfBibitem
\bibitem[Sezen \latin{et~al.}(2015)Sezen, Buchholz, Nefedov, Natzeck, Heissler,
  {Di Valentin}, and W{\"{o}}ll]{Sezen2015a}
Sezen,~H.; Buchholz,~M.; Nefedov,~A.; Natzeck,~C.; Heissler,~S.; {Di
  Valentin},~C.; W{\"{o}}ll,~C. {Probing electrons in TiO$_2$ polaronic trap
  states by IR-absorption: Evidence for the existence of hydrogenic states}.
  \emph{Scientific Reports} \textbf{2015}, \emph{4}, 3808\relax
\mciteBstWouldAddEndPuncttrue
\mciteSetBstMidEndSepPunct{\mcitedefaultmidpunct}
{\mcitedefaultendpunct}{\mcitedefaultseppunct}\relax
\EndOfBibitem
\bibitem[Freytag \latin{et~al.}(2016)Freytag, Corradi, and Imlau]{Freytag2016}
Freytag,~F.; Corradi,~G.; Imlau,~M. {Atomic insight to lattice distortions
  caused by carrier self-trapping in oxide materials}. \emph{Scientific
  Reports} \textbf{2016}, \emph{6}, 36929\relax
\mciteBstWouldAddEndPuncttrue
\mciteSetBstMidEndSepPunct{\mcitedefaultmidpunct}
{\mcitedefaultendpunct}{\mcitedefaultseppunct}\relax
\EndOfBibitem
\bibitem[Crevecoeur and {De Wit}(1970)Crevecoeur, and {De Wit}]{Crevecoeur1970}
Crevecoeur,~C.; {De Wit},~H. {Electrical conductivity of Li doped MnO}.
  \emph{Journal of Physics and Chemistry of Solids} \textbf{1970}, \emph{31},
  783--791\relax
\mciteBstWouldAddEndPuncttrue
\mciteSetBstMidEndSepPunct{\mcitedefaultmidpunct}
{\mcitedefaultendpunct}{\mcitedefaultseppunct}\relax
\EndOfBibitem
\bibitem[Moser \latin{et~al.}(2013)Moser, Moreschini, Ja{\'{c}}imovi{\'{c}},
  Bari{\v{s}}i{\'{c}}, Berger, Magrez, Chang, Kim, Bostwick, Rotenberg,
  Forr{\'{o}}, and Grioni]{Moser2013}
Moser,~S.; Moreschini,~L.; Ja{\'{c}}imovi{\'{c}},~J.;
  Bari{\v{s}}i{\'{c}},~O.~S.; Berger,~H.; Magrez,~A.; Chang,~Y.~J.; Kim,~K.~S.;
  Bostwick,~A.; Rotenberg,~E. \latin{et~al.}  {Tunable polaronic conduction in
  anatase TiO$_2$}. \emph{Physical Review Letters} \textbf{2013}, \emph{110},
  1--5\relax
\mciteBstWouldAddEndPuncttrue
\mciteSetBstMidEndSepPunct{\mcitedefaultmidpunct}
{\mcitedefaultendpunct}{\mcitedefaultseppunct}\relax
\EndOfBibitem
\bibitem[Salje \latin{et~al.}(2005)Salje, Alexandrov, and Liang]{Salje}
Salje,~E.; Alexandrov,~A.~S.; Liang,~W.~Y. \emph{Polarons and Bipolarons in
  High-Tc Superconductors and Related Materials}; Cambridge University Press,
  Cambridge, England, 2005\relax
\mciteBstWouldAddEndPuncttrue
\mciteSetBstMidEndSepPunct{\mcitedefaultmidpunct}
{\mcitedefaultendpunct}{\mcitedefaultseppunct}\relax
\EndOfBibitem
\bibitem[Teresa \latin{et~al.}(1997)Teresa, Ibarra, Algarabel, Ritter,
  Marquina, Blasco, Garcia, del Moral, and Arnold]{Teresa}
Teresa,~J. M.~D.; Ibarra,~M.~R.; Algarabel,~P.~A.; Ritter,~C.; Marquina,~C.;
  Blasco,~J.; Garcia,~J.; del Moral,~A.; Arnold,~Z. Evidence for magnetic
  polarons in the magnetoresistive perovskites. \emph{Nature} \textbf{1997},
  \emph{386}, 256--259\relax
\mciteBstWouldAddEndPuncttrue
\mciteSetBstMidEndSepPunct{\mcitedefaultmidpunct}
{\mcitedefaultendpunct}{\mcitedefaultseppunct}\relax
\EndOfBibitem
\bibitem[R{\o}nnow \latin{et~al.}(2006)R{\o}nnow, Renner, Aeppli, Kimura, and
  Tokura]{Ronnow}
R{\o}nnow,~H.~M.; Renner,~C.; Aeppli,~G.; Kimura,~T.; Tokura,~Y. Polarons and
  confinement of electronic motion to two dimensions in a layered manganite.
  \emph{Nature} \textbf{2006}, \emph{440}, 1025--1028\relax
\mciteBstWouldAddEndPuncttrue
\mciteSetBstMidEndSepPunct{\mcitedefaultmidpunct}
{\mcitedefaultendpunct}{\mcitedefaultseppunct}\relax
\EndOfBibitem
\bibitem[Wang \latin{et~al.}(2014)Wang, Bi, Li, Long, Liu, Lv, Lu, Sun, and
  Liu]{Wang2014}
Wang,~M.; Bi,~C.; Li,~L.; Long,~S.; Liu,~Q.; Lv,~H.; Lu,~N.; Sun,~P.; Liu,~M.
  {Thermoelectric Seebeck effect in oxide-based resistive switching memory}.
  \emph{Nature Communications} \textbf{2014}, \emph{5}, 4598\relax
\mciteBstWouldAddEndPuncttrue
\mciteSetBstMidEndSepPunct{\mcitedefaultmidpunct}
{\mcitedefaultendpunct}{\mcitedefaultseppunct}\relax
\EndOfBibitem
\bibitem[Cortecchia \latin{et~al.}(2017)Cortecchia, Yin, Bruno, Lo, Gurzadyan,
  Mhaisalkar, Br{\'{e}}das, and Soci]{Cortecchia2017}
Cortecchia,~D.; Yin,~J.; Bruno,~A.; Lo,~S.-Z.~A.; Gurzadyan,~G.~G.;
  Mhaisalkar,~S.; Br{\'{e}}das,~J.-L.; Soci,~C. {Polaron self-localization in
  white-light emitting hybrid perovskites}. \emph{Journal of Materials
  Chemistry C} \textbf{2017}, \emph{5}, 2771--2780\relax
\mciteBstWouldAddEndPuncttrue
\mciteSetBstMidEndSepPunct{\mcitedefaultmidpunct}
{\mcitedefaultendpunct}{\mcitedefaultseppunct}\relax
\EndOfBibitem
\bibitem[Linsebigler \latin{et~al.}(1995)Linsebigler, Lu, and
  Yates]{Linsebigler}
Linsebigler,~A.~L.; Lu,~G.; Yates,~J.~T. Photocatalysis on $\mathrm{TiO_2}$
  Surfaces: Principles, Mechanisms, and Selected Results. \emph{Chemical
  Reviews} \textbf{1995}, \emph{95}, 735--758\relax
\mciteBstWouldAddEndPuncttrue
\mciteSetBstMidEndSepPunct{\mcitedefaultmidpunct}
{\mcitedefaultendpunct}{\mcitedefaultseppunct}\relax
\EndOfBibitem
\bibitem[Diebold(2003)]{Diebold2003}
Diebold,~U. {The surface science of titanium dioxide}. \emph{Surface Science
  Reports} \textbf{2003}, \emph{48}, 53--229\relax
\mciteBstWouldAddEndPuncttrue
\mciteSetBstMidEndSepPunct{\mcitedefaultmidpunct}
{\mcitedefaultendpunct}{\mcitedefaultseppunct}\relax
\EndOfBibitem
\bibitem[{Di Valentin} \latin{et~al.}(2006){Di Valentin}, Pacchioni, and
  Selloni]{DiValentin2006}
{Di Valentin},~C.; Pacchioni,~G.; Selloni,~A. {Electronic Structure of Defect
  States in Hydroxylated and Reduced Rutile TiO$_2$(110) Surfaces}.
  \emph{Physical Review Letters} \textbf{2006}, \emph{97}, 166803\relax
\mciteBstWouldAddEndPuncttrue
\mciteSetBstMidEndSepPunct{\mcitedefaultmidpunct}
{\mcitedefaultendpunct}{\mcitedefaultseppunct}\relax
\EndOfBibitem
\bibitem[De{\'{a}}k \latin{et~al.}(2012)De{\'{a}}k, Aradi, and
  Frauenheim]{Deak2012b}
De{\'{a}}k,~P.; Aradi,~B.; Frauenheim,~T. {Quantitative theory of the oxygen
  vacancy and carrier self-trapping in bulk TiO$_2$}. \emph{Physical Review B -
  Condensed Matter and Materials Physics} \textbf{2012}, \emph{86}, 1--8\relax
\mciteBstWouldAddEndPuncttrue
\mciteSetBstMidEndSepPunct{\mcitedefaultmidpunct}
{\mcitedefaultendpunct}{\mcitedefaultseppunct}\relax
\EndOfBibitem
\bibitem[Berger \latin{et~al.}(2015)Berger, Oberhofer, and Reuter]{Berger2015b}
Berger,~D.; Oberhofer,~H.; Reuter,~K. {First-principles embedded-cluster
  calculations of the neutral and charged oxygen vacancy at the rutile
  TiO$_2$(110) surface}. \emph{Physical Review B - Condensed Matter and
  Materials Physics} \textbf{2015}, \emph{92}, 1--11\relax
\mciteBstWouldAddEndPuncttrue
\mciteSetBstMidEndSepPunct{\mcitedefaultmidpunct}
{\mcitedefaultendpunct}{\mcitedefaultseppunct}\relax
\EndOfBibitem
\bibitem[Yan \latin{et~al.}(2015)Yan, Elenewski, Jiang, and Chen]{Yan2015}
Yan,~L.; Elenewski,~J.~E.; Jiang,~W.; Chen,~H. {Computational modeling of
  self-trapped electrons in rutile TiO$_2$}. \emph{Phys. Chem. Chem. Phys.}
  \textbf{2015}, \emph{17}, 29949--29957\relax
\mciteBstWouldAddEndPuncttrue
\mciteSetBstMidEndSepPunct{\mcitedefaultmidpunct}
{\mcitedefaultendpunct}{\mcitedefaultseppunct}\relax
\EndOfBibitem
\bibitem[Kr{\"{u}}ger \latin{et~al.}(2008)Kr{\"{u}}ger, Bourgeois, Domenichini,
  Magnan, Chandesris, {Le F{\`{e}}vre}, Flank, Jupille, Floreano, Cossaro,
  Verdini, and Morgante]{Kruger2008}
Kr{\"{u}}ger,~P.; Bourgeois,~S.; Domenichini,~B.; Magnan,~H.; Chandesris,~D.;
  {Le F{\`{e}}vre},~P.; Flank,~A.~M.; Jupille,~J.; Floreano,~L.; Cossaro,~A.
  \latin{et~al.}  {Defect states at the TiO$_2$(110) surface probed by resonant
  photoelectron diffraction}. \emph{Physical Review Letters} \textbf{2008},
  \emph{100}, 2--5\relax
\mciteBstWouldAddEndPuncttrue
\mciteSetBstMidEndSepPunct{\mcitedefaultmidpunct}
{\mcitedefaultendpunct}{\mcitedefaultseppunct}\relax
\EndOfBibitem
\bibitem[Yang \latin{et~al.}(2013)Yang, Brant, Giles, and
  Halliburton]{Yang2013}
Yang,~S.; Brant,~A.~T.; Giles,~N.~C.; Halliburton,~L.~E. {Intrinsic small
  polarons in rutile TiO$_2$}. \emph{Physical Review B} \textbf{2013},
  \emph{87}, 125201\relax
\mciteBstWouldAddEndPuncttrue
\mciteSetBstMidEndSepPunct{\mcitedefaultmidpunct}
{\mcitedefaultendpunct}{\mcitedefaultseppunct}\relax
\EndOfBibitem
\bibitem[Setvin \latin{et~al.}(2014)Setvin, Franchini, Hao, Schmid, Janotti,
  Kaltak, {Van De Walle}, Kresse, and Diebold]{Setvin2014}
Setvin,~M.; Franchini,~C.; Hao,~X.; Schmid,~M.; Janotti,~A.; Kaltak,~M.; {Van
  De Walle},~C.~G.; Kresse,~G.; Diebold,~U. {Direct view at excess electrons in
  TiO$_2$ rutile and anatase}. \emph{Physical Review Letters} \textbf{2014},
  \emph{113}, 1--5\relax
\mciteBstWouldAddEndPuncttrue
\mciteSetBstMidEndSepPunct{\mcitedefaultmidpunct}
{\mcitedefaultendpunct}{\mcitedefaultseppunct}\relax
\EndOfBibitem
\bibitem[Shibuya \latin{et~al.}(2014)Shibuya, Yasuoka, Mirbt, and
  Sanyal]{Shibuya2014}
Shibuya,~T.; Yasuoka,~K.; Mirbt,~S.; Sanyal,~B. {Bipolaron Formation Induced by
  Oxygen Vacancy at Rutile TiO$_2$ (110) Surfaces}. \emph{The Journal of
  Physical Chemistry C} \textbf{2014}, \emph{118}, 9429--9435\relax
\mciteBstWouldAddEndPuncttrue
\mciteSetBstMidEndSepPunct{\mcitedefaultmidpunct}
{\mcitedefaultendpunct}{\mcitedefaultseppunct}\relax
\EndOfBibitem
\bibitem[Mao \latin{et~al.}(2013)Mao, Lang, Wang, Hao, Wen, Ren, Dai, Zhou,
  Liu, and Yang]{Mao2013}
Mao,~X.; Lang,~X.; Wang,~Z.; Hao,~Q.; Wen,~B.; Ren,~Z.; Dai,~D.; Zhou,~C.;
  Liu,~L.-M.; Yang,~X. {Band-Gap States of TiO$_2$(110): Major Contribution
  from Surface Defects}. \emph{The Journal of Physical Chemistry Letters}
  \textbf{2013}, \emph{4}, 3839--3844\relax
\mciteBstWouldAddEndPuncttrue
\mciteSetBstMidEndSepPunct{\mcitedefaultmidpunct}
{\mcitedefaultendpunct}{\mcitedefaultseppunct}\relax
\EndOfBibitem
\bibitem[Finazzi \latin{et~al.}(2009)Finazzi, Valentin, Pacchioni, {Di
  Valentin}, and Pacchioni]{Finazzi2009}
Finazzi,~E.; Valentin,~C.~D.; Pacchioni,~G.; {Di Valentin},~C.; Pacchioni,~G.
  {Nature of Ti Interstitials in Reduced Bulk Anatase and Rutile TiO$_2$}.
  \emph{The Journal of Physical Chemistry C} \textbf{2009}, \emph{113},
  3382--3385\relax
\mciteBstWouldAddEndPuncttrue
\mciteSetBstMidEndSepPunct{\mcitedefaultmidpunct}
{\mcitedefaultendpunct}{\mcitedefaultseppunct}\relax
\EndOfBibitem
\bibitem[Deskins \latin{et~al.}(2011)Deskins, Rousseau, and
  Dupuis]{Deskins2011}
Deskins,~N.~A.; Rousseau,~R.; Dupuis,~M. {Distribution of Ti3+Surface Sites in
  Reduced TiO$_2$}. \emph{Journal of Physical Chemistry C} \textbf{2011},
  \emph{115}, 7562--7572\relax
\mciteBstWouldAddEndPuncttrue
\mciteSetBstMidEndSepPunct{\mcitedefaultmidpunct}
{\mcitedefaultendpunct}{\mcitedefaultseppunct}\relax
\EndOfBibitem
\bibitem[Cao \latin{et~al.}(2017)Cao, Yu, Qi, Huang, Wang, Xu, Hu, and
  Yan]{Cao2017a}
Cao,~Y.; Yu,~M.; Qi,~S.; Huang,~S.; Wang,~T.; Xu,~M.; Hu,~S.; Yan,~S.
  {Scenarios of polaron-involved molecular adsorption on reduced TiO$_2$(110)
  surfaces}. \emph{Scientific Reports} \textbf{2017}, \emph{7}, 6148\relax
\mciteBstWouldAddEndPuncttrue
\mciteSetBstMidEndSepPunct{\mcitedefaultmidpunct}
{\mcitedefaultendpunct}{\mcitedefaultseppunct}\relax
\EndOfBibitem
\bibitem[Reticcioli \latin{et~al.}(2017)Reticcioli, Setvin, Hao, Flauger,
  Kresse, Schmid, Diebold, and Franchini]{Reticcioli2017c}
Reticcioli,~M.; Setvin,~M.; Hao,~X.; Flauger,~P.; Kresse,~G.; Schmid,~M.;
  Diebold,~U.; Franchini,~C. {Polaron-Driven Surface Reconstructions}.
  \emph{Physical Review X} \textbf{2017}, \emph{7}, 031053\relax
\mciteBstWouldAddEndPuncttrue
\mciteSetBstMidEndSepPunct{\mcitedefaultmidpunct}
{\mcitedefaultendpunct}{\mcitedefaultseppunct}\relax
\EndOfBibitem
\bibitem[Carneiro \latin{et~al.}(2017)Carneiro, Cushing, Liu, Su, Yang,
  Alivisatos, and Leone]{Carneiro2017}
Carneiro,~L.~M.; Cushing,~S.~K.; Liu,~C.; Su,~Y.; Yang,~P.; Alivisatos,~A.~P.;
  Leone,~S.~R. {Excitation-wavelength-dependent small polaron trapping of
  photoexcited carriers in $\alpha$-Fe2O3}. \emph{Nature Materials}
  \textbf{2017}, \emph{16}, 819--825\relax
\mciteBstWouldAddEndPuncttrue
\mciteSetBstMidEndSepPunct{\mcitedefaultmidpunct}
{\mcitedefaultendpunct}{\mcitedefaultseppunct}\relax
\EndOfBibitem
\bibitem[Setvin \latin{et~al.}(2014)Setvin, Hao, Daniel, Pavelec, Novotny,
  Parkinson, Schmid, Kresse, Franchini, and Diebold]{Setvin2014b}
Setvin,~M.; Hao,~X.; Daniel,~B.; Pavelec,~J.; Novotny,~Z.; Parkinson,~G.~S.;
  Schmid,~M.; Kresse,~G.; Franchini,~C.; Diebold,~U. {Charge Trapping at the
  Step Edges of TiO$_2$ Anatase (101)}. \emph{Angewandte Chemie International
  Edition} \textbf{2014}, \emph{53}, 4714--4716\relax
\mciteBstWouldAddEndPuncttrue
\mciteSetBstMidEndSepPunct{\mcitedefaultmidpunct}
{\mcitedefaultendpunct}{\mcitedefaultseppunct}\relax
\EndOfBibitem
\bibitem[Kowalski \latin{et~al.}(2010)Kowalski, Camellone, Nair, Meyer, and
  Marx]{Kowalski2010}
Kowalski,~P.~M.; Camellone,~M.~F.; Nair,~N.~N.; Meyer,~B.; Marx,~D. {Charge
  localization dynamics induced by oxygen vacancies on the TiO$_2$(110)
  surface}. \emph{Physical Review Letters} \textbf{2010}, \emph{105},
  5--8\relax
\mciteBstWouldAddEndPuncttrue
\mciteSetBstMidEndSepPunct{\mcitedefaultmidpunct}
{\mcitedefaultendpunct}{\mcitedefaultseppunct}\relax
\EndOfBibitem
\bibitem[Deskins and Dupuis(2007)Deskins, and Dupuis]{Deskins2007}
Deskins,~N.~A.; Dupuis,~M. {Electron transport via polaron hopping in bulk Ti
  O2: A density functional theory characterization}. \emph{Physical Review B -
  Condensed Matter and Materials Physics} \textbf{2007}, \emph{75}, 1--10\relax
\mciteBstWouldAddEndPuncttrue
\mciteSetBstMidEndSepPunct{\mcitedefaultmidpunct}
{\mcitedefaultendpunct}{\mcitedefaultseppunct}\relax
\EndOfBibitem
\bibitem[Kresse and Furthm{\"{u}}ller(1996)Kresse, and
  Furthm{\"{u}}ller]{Kresse1996a}
Kresse,~G.; Furthm{\"{u}}ller,~J. {Efficient iterative schemes for ab initio
  total-energy calculations using a plane-wave basis set}. \emph{Physical
  Review B} \textbf{1996}, \emph{54}, 11169--11186\relax
\mciteBstWouldAddEndPuncttrue
\mciteSetBstMidEndSepPunct{\mcitedefaultmidpunct}
{\mcitedefaultendpunct}{\mcitedefaultseppunct}\relax
\EndOfBibitem
\bibitem[Kresse and Furthm{\"{u}}ller(1996)Kresse, and
  Furthm{\"{u}}ller]{Kresse1996}
Kresse,~G.; Furthm{\"{u}}ller,~J. {Efficiency of ab-initio total energy
  calculations for metals and semiconductors using a plane-wave basis set}.
  \emph{Computational Materials Science} \textbf{1996}, \emph{6}, 15--50\relax
\mciteBstWouldAddEndPuncttrue
\mciteSetBstMidEndSepPunct{\mcitedefaultmidpunct}
{\mcitedefaultendpunct}{\mcitedefaultseppunct}\relax
\EndOfBibitem
\bibitem[Perdew \latin{et~al.}(1996)Perdew, Burke, and Ernzerhof]{Perdew1996}
Perdew,~J.~P.; Burke,~K.; Ernzerhof,~M. {Generalized Gradient Approximation
  Made Simple}. \emph{Physical Review Letters} \textbf{1996}, \emph{77},
  3865--3868\relax
\mciteBstWouldAddEndPuncttrue
\mciteSetBstMidEndSepPunct{\mcitedefaultmidpunct}
{\mcitedefaultendpunct}{\mcitedefaultseppunct}\relax
\EndOfBibitem
\bibitem[Dudarev \latin{et~al.}(1998)Dudarev, Botton, Savrasov, Humphreys, and
  Sutton]{Dudarev1998}
Dudarev,~S.~L.; Botton,~G.~A.; Savrasov,~S.~Y.; Humphreys,~C.~J.; Sutton,~a.~P.
  {Electron-energy-loss spectra and the structural stability of nickel oxide:
  An LSDA+U study}. \emph{Physical Review B} \textbf{1998}, \emph{57},
  1505--1509\relax
\mciteBstWouldAddEndPuncttrue
\mciteSetBstMidEndSepPunct{\mcitedefaultmidpunct}
{\mcitedefaultendpunct}{\mcitedefaultseppunct}\relax
\EndOfBibitem
\bibitem[Wang \latin{et~al.}(2017)Wang, Brock, Matt, and Bevan]{Wang2017}
Wang,~Z.; Brock,~C.; Matt,~A.; Bevan,~K.~H. {Implications of the DFT+U method
  on polaron properties in energy materials}. \emph{Physical Review B}
  \textbf{2017}, \emph{96}, 1--13\relax
\mciteBstWouldAddEndPuncttrue
\mciteSetBstMidEndSepPunct{\mcitedefaultmidpunct}
{\mcitedefaultendpunct}{\mcitedefaultseppunct}\relax
\EndOfBibitem
\bibitem[Onishi and Iwasawa(1994)Onishi, and Iwasawa]{Onishi1994}
Onishi,~H.; Iwasawa,~Y. {Reconstruction of TiO$_2$(110) surface: STM study with
  atomic-scale resolution}. \emph{Surface Science} \textbf{1994}, \emph{313},
  L783--L789\relax
\mciteBstWouldAddEndPuncttrue
\mciteSetBstMidEndSepPunct{\mcitedefaultmidpunct}
{\mcitedefaultendpunct}{\mcitedefaultseppunct}\relax
\EndOfBibitem
\bibitem[Wang \latin{et~al.}(2014)Wang, Oganov, Zhu, and Zhou]{Wang2014a}
Wang,~Q.; Oganov,~A.~R.; Zhu,~Q.; Zhou,~X.~F. {New reconstructions of the (110)
  surface of rutile TiO$_2$ predicted by an evolutionary method}.
  \emph{Physical Review Letters} \textbf{2014}, \emph{113}, 1--5\relax
\mciteBstWouldAddEndPuncttrue
\mciteSetBstMidEndSepPunct{\mcitedefaultmidpunct}
{\mcitedefaultendpunct}{\mcitedefaultseppunct}\relax
\EndOfBibitem
\bibitem[Mochizuki \latin{et~al.}(2016)Mochizuki, Ariga, Fukaya, Wada, Maekawa,
  Kawasuso, Shidara, Asakura, and Hyodo]{Mochizuki2016}
Mochizuki,~I.; Ariga,~H.; Fukaya,~Y.; Wada,~K.; Maekawa,~M.; Kawasuso,~A.;
  Shidara,~T.; Asakura,~K.; Hyodo,~T. {Structure determination of the
  rutile-TiO$_2$(110)-(1$\times$2) surface using total-reflection high-energy
  positron diffraction (TRHEPD)}. \emph{Phys. Chem. Chem. Phys.} \textbf{2016},
  \emph{18}, 7085--7092\relax
\mciteBstWouldAddEndPuncttrue
\mciteSetBstMidEndSepPunct{\mcitedefaultmidpunct}
{\mcitedefaultendpunct}{\mcitedefaultseppunct}\relax
\EndOfBibitem
\bibitem[Cheng and Selloni(2009)Cheng, and Selloni]{Cheng2009}
Cheng,~H.; Selloni,~A. {Surface and subsurface oxygen vacancies in anatase
  TiO$_2$ and differences with rutile}. \emph{Physical Review B} \textbf{2009},
  \emph{79}, 2--5\relax
\mciteBstWouldAddEndPuncttrue
\mciteSetBstMidEndSepPunct{\mcitedefaultmidpunct}
{\mcitedefaultendpunct}{\mcitedefaultseppunct}\relax
\EndOfBibitem
\bibitem[Li \latin{et~al.}()Li, Guo, and Robertson]{Li2015}
Li,~H.; Guo,~Y.; Robertson,~J. {Calculation of TiO$_2$} Surface and Subsurface
  Oxygen Vacancy by the Screened Exchange Functional. \emph{Journal of Physical
  Chemistry C} 18160--18166\relax
\mciteBstWouldAddEndPuncttrue
\mciteSetBstMidEndSepPunct{\mcitedefaultmidpunct}
{\mcitedefaultendpunct}{\mcitedefaultseppunct}\relax
\EndOfBibitem
\bibitem[Kresse and Hafner(1993)Kresse, and Hafner]{Kresse1993}
Kresse,~G.; Hafner,~J. {Ab initio molecular dynamics for liquid metals}.
  \emph{Physical Review B} \textbf{1993}, \emph{47}, 558--561\relax
\mciteBstWouldAddEndPuncttrue
\mciteSetBstMidEndSepPunct{\mcitedefaultmidpunct}
{\mcitedefaultendpunct}{\mcitedefaultseppunct}\relax
\EndOfBibitem
\bibitem[Hao \latin{et~al.}(2015)Hao, Wang, Schmid, Diebold, and
  Franchini]{Hao2015a}
Hao,~X.; Wang,~Z.; Schmid,~M.; Diebold,~U.; Franchini,~C. {Coexistence of
  trapped and free excess electrons in SrTiO$_3$}. \emph{Physical Review B}
  \textbf{2015}, \emph{91}, 085204\relax
\mciteBstWouldAddEndPuncttrue
\mciteSetBstMidEndSepPunct{\mcitedefaultmidpunct}
{\mcitedefaultendpunct}{\mcitedefaultseppunct}\relax
\EndOfBibitem
\bibitem[Papageorgiou \latin{et~al.}(2010)Papageorgiou, Beglitis, Pang,
  Teobaldi, Cabailh, Chen, Fisher, Hofer, and Thornton]{Papageorgiou2010a}
Papageorgiou,~A.~C.; Beglitis,~N.~S.; Pang,~C.~L.; Teobaldi,~G.; Cabailh,~G.;
  Chen,~Q.; Fisher,~A.~J.; Hofer,~W.~A.; Thornton,~G. {Electron traps and their
  effect on the surface chemistry of TiO$_2$(110)}. \emph{Proceedings of the
  National Academy of Sciences} \textbf{2010}, \emph{107}, 2391--2396\relax
\mciteBstWouldAddEndPuncttrue
\mciteSetBstMidEndSepPunct{\mcitedefaultmidpunct}
{\mcitedefaultendpunct}{\mcitedefaultseppunct}\relax
\EndOfBibitem
\bibitem[Deskins \latin{et~al.}(2009)Deskins, Rousseau, and
  Dupuis]{Deskins2009}
Deskins,~N.~A.; Rousseau,~R.; Dupuis,~M. {Localized electronic states from
  surface hydroxyls and polarons in TiO$_2$(110)}. \emph{Journal of Physical
  Chemistry C} \textbf{2009}, \emph{113}, 14583--14586\relax
\mciteBstWouldAddEndPuncttrue
\mciteSetBstMidEndSepPunct{\mcitedefaultmidpunct}
{\mcitedefaultendpunct}{\mcitedefaultseppunct}\relax
\EndOfBibitem
\bibitem[Meredig \latin{et~al.}(2010)Meredig, Thompson, Hansen, Wolverton, and
  van~de Walle]{Meredig2010a}
Meredig,~B.; Thompson,~A.; Hansen,~H.~A.; Wolverton,~C.; van~de Walle,~A.
  {Method for locating low-energy solutions within DFT+$U$}. \emph{Physical
  Review B} \textbf{2010}, \emph{82}, 195128\relax
\mciteBstWouldAddEndPuncttrue
\mciteSetBstMidEndSepPunct{\mcitedefaultmidpunct}
{\mcitedefaultendpunct}{\mcitedefaultseppunct}\relax
\EndOfBibitem
\bibitem[Morgan and Watson(2009)Morgan, and Watson]{Morgan2009}
Morgan,~B.~J.; Watson,~G.~W. {Polaronic trapping of electrons and holes by
  native defects in anatase TiO$_2$}. \emph{Physical Review B - Condensed
  Matter and Materials Physics} \textbf{2009}, \emph{80}, 2--5\relax
\mciteBstWouldAddEndPuncttrue
\mciteSetBstMidEndSepPunct{\mcitedefaultmidpunct}
{\mcitedefaultendpunct}{\mcitedefaultseppunct}\relax
\EndOfBibitem
\bibitem[Janotti \latin{et~al.}(2013)Janotti, Franchini, Varley, Kresse, and
  {Van de Walle}]{Janotti2013a}
Janotti,~A.; Franchini,~C.; Varley,~J.~B.; Kresse,~G.; {Van de Walle},~C.~G.
  {Dual behavior of excess electrons in rutile TiO$_2$}. \emph{physica status
  solidi (RRL) - Rapid Research Letters} \textbf{2013}, \emph{7},
  199--203\relax
\mciteBstWouldAddEndPuncttrue
\mciteSetBstMidEndSepPunct{\mcitedefaultmidpunct}
{\mcitedefaultendpunct}{\mcitedefaultseppunct}\relax
\EndOfBibitem
\bibitem[{Amore Bonapasta} \latin{et~al.}(2009){Amore Bonapasta}, Filippone,
  Mattioli, and Alippi]{AmoreBonapasta2009}
{Amore Bonapasta},~A.; Filippone,~F.; Mattioli,~G.; Alippi,~P. {Oxygen
  vacancies and OH species in rutile and anatase TiO$_2$ polymorphs}.
  \emph{Catalysis Today} \textbf{2009}, \emph{144}, 177--182\relax
\mciteBstWouldAddEndPuncttrue
\mciteSetBstMidEndSepPunct{\mcitedefaultmidpunct}
{\mcitedefaultendpunct}{\mcitedefaultseppunct}\relax
\EndOfBibitem
\bibitem[Yan \latin{et~al.}(2015)Yan, Elenewski, Jiang, and Chen]{Yan2015a}
Yan,~L.; Elenewski,~J.~E.; Jiang,~W.; Chen,~H. {Computational modeling of
  self-trapped electrons in rutile TiO$_2$}. \emph{Physical Chemistry Chemical
  Physics} \textbf{2015}, \emph{17}, 29949--29957\relax
\mciteBstWouldAddEndPuncttrue
\mciteSetBstMidEndSepPunct{\mcitedefaultmidpunct}
{\mcitedefaultendpunct}{\mcitedefaultseppunct}\relax
\EndOfBibitem
\bibitem[Yim \latin{et~al.}(2016)Yim, Watkins, Wolf, Pang, Hermansson, and
  Thornton]{Yim2016a}
Yim,~C.~M.; Watkins,~M.~B.; Wolf,~M.~J.; Pang,~C.~L.; Hermansson,~K.;
  Thornton,~G. {Engineering Polarons at a Metal Oxide Surface}. \emph{Physical
  Review Letters} \textbf{2016}, \emph{117}, 1--16\relax
\mciteBstWouldAddEndPuncttrue
\mciteSetBstMidEndSepPunct{\mcitedefaultmidpunct}
{\mcitedefaultendpunct}{\mcitedefaultseppunct}\relax
\EndOfBibitem
\bibitem[Shibuya \latin{et~al.}(2012)Shibuya, Yasuoka, Mirbt, and
  Sanyal]{Shibuya2012}
Shibuya,~T.; Yasuoka,~K.; Mirbt,~S.; Sanyal,~B. {A systematic study of polarons
  due to oxygen vacancy formation at the rutile TiO$_2$(110) surface by GGA + U
  and HSE06 methods}. \emph{Journal of Physics: Condensed Matter}
  \textbf{2012}, \emph{24}, 435504\relax
\mciteBstWouldAddEndPuncttrue
\mciteSetBstMidEndSepPunct{\mcitedefaultmidpunct}
{\mcitedefaultendpunct}{\mcitedefaultseppunct}\relax
\EndOfBibitem
\bibitem[Calzado \latin{et~al.}(2008)Calzado, Hern{\'{a}}ndez, and
  Sanz]{Calzado2008}
Calzado,~C.~J.; Hern{\'{a}}ndez,~N.~C.; Sanz,~J.~F. {Effect of on-site Coulomb
  repulsion term U on the band-gap states of the reduced rutile (110) Ti O2
  surface}. \emph{Physical Review B - Condensed Matter and Materials Physics}
  \textbf{2008}, \emph{77}, 1--10\relax
\mciteBstWouldAddEndPuncttrue
\mciteSetBstMidEndSepPunct{\mcitedefaultmidpunct}
{\mcitedefaultendpunct}{\mcitedefaultseppunct}\relax
\EndOfBibitem
\bibitem[Moses \latin{et~al.}(2016)Moses, Janotti, Franchini, Kresse, and {Van
  de Walle}]{Moses2016}
Moses,~P.~G.; Janotti,~A.; Franchini,~C.; Kresse,~G.; {Van de Walle},~C.~G.
  {Donor defects and small polarons on the TiO$_2$(110) surface}. \emph{Journal
  of Applied Physics} \textbf{2016}, \emph{119}, 181503\relax
\mciteBstWouldAddEndPuncttrue
\mciteSetBstMidEndSepPunct{\mcitedefaultmidpunct}
{\mcitedefaultendpunct}{\mcitedefaultseppunct}\relax
\EndOfBibitem
\bibitem[Mackrodt \latin{et~al.}(1997)Mackrodt, Simson, and
  Harrison]{Mackrodt1997}
Mackrodt,~W.; Simson,~E.-a.; Harrison,~N. {An ab initio Hartree-Fock study of
  the electron-excess gap states in oxygen-deficient rutile TiO2}.
  \emph{Surface Science} \textbf{1997}, \emph{384}, 192--200\relax
\mciteBstWouldAddEndPuncttrue
\mciteSetBstMidEndSepPunct{\mcitedefaultmidpunct}
{\mcitedefaultendpunct}{\mcitedefaultseppunct}\relax
\EndOfBibitem
\bibitem[Lakkis \latin{et~al.}(1976)Lakkis, Schlenker, Chakraverty, Buder, and
  Marezio]{Lakkis1976}
Lakkis,~S.; Schlenker,~C.; Chakraverty,~B.~K.; Buder,~R.; Marezio,~M.
  {Metal-insulator transitions in Ti$_4$O$_7$ single crystals: Crystal
  characterization, specific heat, and electron paramagnetic resonance}.
  \emph{Physical Review B} \textbf{1976}, \emph{14}, 1429--1440\relax
\mciteBstWouldAddEndPuncttrue
\mciteSetBstMidEndSepPunct{\mcitedefaultmidpunct}
{\mcitedefaultendpunct}{\mcitedefaultseppunct}\relax
\EndOfBibitem
\bibitem[Bondarenko \latin{et~al.}(2015)Bondarenko, Eriksson, and
  Skorodumova]{Bondarenko2015}
Bondarenko,~N.; Eriksson,~O.; Skorodumova,~N.~V. {Polaron mobility in
  oxygen-deficient and lithium-doped tungsten trioxide}. \emph{Physical Review
  B} \textbf{2015}, \emph{92}, 165119\relax
\mciteBstWouldAddEndPuncttrue
\mciteSetBstMidEndSepPunct{\mcitedefaultmidpunct}
{\mcitedefaultendpunct}{\mcitedefaultseppunct}\relax
\EndOfBibitem
\end{mcitethebibliography}

\end{document}